\documentclass[12pt,twoside]{article} 
\usepackage[latin1]{inputenc}
\usepackage{graphicx}
\usepackage{latexsym}
\usepackage{amssymb}
\usepackage{epsfig}
\usepackage{hyperref}
\usepackage{amsmath}
\usepackage{natbib}
\usepackage{float}

\newcommand{\Var}{\mbox{Var}}
\def\eps{\varepsilon}

\newcommand{\nb}{\nonumber}

\newcommand{\ds}{\displaystyle}

\newcommand{\ns}{\lfloor ns\rfloor}

\def\en{\mathbb{N}}
\def\zet{\mathbb{Z}}
\def\er{\mathbb{R}}
\def\ge{\mathbb{G}}
\def\ka{\mathbb{K}}

\def\cn{\mathcal{N}}

\def\cd{\mathcal{D}}
\def\cf{\mathcal{F}}
\def\e{\varepsilon}
\def\he{\hat{\varepsilon}}
\def\be{\bar{\epsilon}_n}

\def\sjn{\sum_{j=1}^n}
\def\op{o_P(\frac 1{\sqrt n})}

\def\beq{\begin{eqnarray*}}
\def\eeq{\end{eqnarray*}}
\def\farc{\frac}
\def\sjns{\sum_{j=1}^{\lfloor ns\rfloor}}
\def\sjnsm{\sum_{j=\lfloor ns\rfloor+1}^n}
\def\fns{\frac{\lfloor ns\rfloor}n}

\def\fnsk{\frac{\sum_{k=1}^{\lfloor ns\rfloor}w_{nk}}{n}}
\def\fnskm{\frac{\sum_{k=\lfloor ns\rfloor+1}^nw_{nk}}{n}}
\def\bK{\bar{K}}
\def\bM{\bar{M}}

\def\ns{\lfloor ns\rfloor}

\def\nt{\lfloor n\theta_0\rfloor}

\def\sjnsh{\sum_{j=1}^{\lfloor ns_h\rfloor}}

\def\nn{\nonumber}

\begin{document}

\title{\bf 
Testing for a change of the innovation distribution in nonparametric autoregression -- the sequential empirical process approach 
}

\author{ Leonie Selk\footnote{corresponding author; e-mail leonie.selk@math.uni-hamburg.de} \, and Natalie Neumeyer\\
University of Hamburg, Department of Mathematics \\
Bundesstrasse 55, 20146 Hamburg, Germany\\
}

\date{January 24, 2012}

\maketitle

\newtheorem{theo}{Theorem}[section]
\newtheorem{lemma}[theo]{Lemma}
\newtheorem{cor}[theo]{Corollary}
\newtheorem{rem}[theo]{Remark}
\newtheorem{prop}[theo]{Proposition}
\newtheorem{defin}[theo]{Definition}
\newtheorem{example}[theo]{Example}

\begin{abstract} 
We consider a nonparametric autoregression model under conditional heteroscedasticity with the aim to test whether the innovation distribution changes in time. 
To this end we develop an asymptotic expansion for the sequential empirical process of nonparametrically estimated innovations (residuals). We suggest a Kolmogorov-Smirnov statistic based on the difference of the estimated innovation distributions built from the first $\ns$ and the last $n-\ns$ residuals, respectively ($0 \le s \le 1$). Weak convergence of the underlying stochastic process to a Gaussian process is proved under the null hypothesis of no change point. The result implies that the test is asymptotically distribution-free. Consistency against fixed alternatives is shown. 
The small sample performances of the proposed test is investigated in a simulation study and the test is applied to data examples.  
\end{abstract}

Running title: Testing for changes in nonparametric autoregression



\section{Introduction}
\def\theequation{1.\arabic{equation}}
\setcounter{equation}{0}

Assume we have observed a time series that can be modelled via an autoregression model, possibly with conditional heteroscedasticity. We aim at testing for a change point in the innovation distribution. 
Tests for change points in the distribution of time series data have received a lot of attention in mathematical statistics; see Picard (1985), Giraitis, Leipus \& Surgailis (1996), Horv\'ath, Kokoszka \& TeyssiÞre (2001), Inoue (2001), Boldin (2002), Lee \& Na (2004), Hu\v{s}kovß, Prß\v{s}kovß \& Steinebach (2007),  Hu\v{s}kovß, Kirch, Prß\v{s}kovß \& Steinebach (2008), among others. Recently, an online-monitoring procedure to detect changes in the innovation distribution of linear autoregressive models was developed by Hlßvka, Hu\v{s}kovß, Kirch \& Meintanis (2012).
Those tests have applications in different areas, e.\,g.\ finance, climate science and medicine. For instance financial time series are tested for changes in the volatility or return (see e.\,g. Andreou \& Ghysels (2009)) or, for climate control reasons, the annual water flow of rivers are tested for changes (see Hu\v{s}kovß \& Antoch (2003)).
 
Classical tests for change points in the distribution of independent data are often based on the difference of empirical distributions of the first $\ns$ and the last $n-\ns$ observations, respectively ($0 \le s \le 1$). To derive asymptotic properties of the test sequential empirical processes are considered; see Shorack \& Wellner (1986, p.\ 131) and also Cs\"org\"o, Horvßth \& Szyszkowicz (1997). 
Those methods for independent data have been transferred to test for change points in the innovation distribution of parametric time series models. Sequential empirical processes based on estimated residuals  and corresponding change point tests were suggested by Bai (1994) for ARMA-models, by Koul (1996) in the context of nonlinear time series and by Ling (1998) for nonstationary autoregressive models. Those articles are the ones most similar in spirit to the paper at hand. However, we do not assume any parametric model for either the autoregression function, nor for the conditonal variance function, but use nonparametric kernel estimation methods. 
The (non-sequential) empirical process of residuals in a nonparametric homoscedastic autoregressive time series model was considered by M³ller, Schick \& Wefelmeyer (2009) who prove an asymptotic expansion. Moreover, residual empirical processes play an important role in the test for multiplicative structure in  a nonparametric heteroscedastic time series regression model by Dette, Pardo-Fernßndez \& Van Keilegom (2009).
On the other hand our approach is similar in spirit to Neumeyer \& Van Keilegom (2009) who consider change point tests for the error distribution in nonparametric regression models with independent observations. 
However, in comparison to the latter three articles the methods of proof in the paper at hand require 
considerably more technical effort because both the time series structure of the data and the additional index $s\in [0,1]$ in the stochastic process have to be taken into account.

We prove an asymptotic expansion for the sequential empirical process of residuals and prove weak convergence of the scaled and centered process to a Gaussian process. It can be seen from those results that the nonparametric estimation of the autoregression and variance function decisively changes the asymptotic behaviour in comparison to the case where innovations would be known. The asymptotic expansion of the sequential process is then used to show that nevertheless the Kolmogorov-Smirnov test for a change point as described above is asymptotically distribution-free. 
As a by-product of our proofs we obtain results on uniform rates of convergence of kernel estimators (see Lemma \ref{lemma3.3} in the appendix). Those are similar in spirit to results derived by Hansen (2008), but in contrast we avoid the stationarity assumption. Only some stabilization of the mean of innovation densities is needed (see assumption (F')), which allows us to apply the results to prove consistency of the test under the existence of a change point. We moreover present a simulation study which shows good approximations of the asymptotic level as well as good power properties of the test under the example models considered. As data applications we consider two financial time series, namely the quarterly GNP of the USA and the S\&P 500 index. 

The paper is organized as follows. In section 2 we present the model, the nonparametric curve estimators and the stochastic process used for the change point test. In section 3 we list  technical assumptions and present the asymptotic results for the sequential empirical process as well as for the process used for the change point test under the null hypothesis of no change point. Asymptotic results under fixed alternatives are presented in section 4. Section 5 is concerned with a homoscedastic modification of the model. In section 6 we present simulation results and consider the data examples.  Section 7 concludes the paper, whereas all proofs are given in the appendix.

\section{Model, hypotheses and test statistic}\label{sec-null}
\def\theequation{2.\arabic{equation}}
\setcounter{equation}{0}

Let $(X_j)_{j\in\zet}$ be a real valued stochastic process following the heteroscedastic autoregressive model of order one, 
\begin{description}
\item [(AR)]  
\qquad\qquad $X_j=m(X_{j-1})+\sigma(X_{j-1})\e_j$,\\
where the innovations $\e_j$, $j\in\zet$, are independent with $E[\e_j]=0$ and $E[\e_j^2]=1$ $\forall j$ and $\e_j$ is independent of the past $X_k$, $k\leq j-1$, $\forall j$.
\end{description}
Assume we have observed $X_0,\ldots,X_n$ and our aim is to test for a change point in the innovation distribution.  
Thus we formulate the null hypothesis as
\[H_0:\qquad\qquad \e_1,\ldots,\e_n\sim F\]
(with $F$ unknown)
while the fixed alternative has the form
\[H_1:\qquad\qquad\exists\ \theta_0\in(0,1):\quad\e_1,\ldots,\e_{\lfloor n\theta_0\rfloor}\sim F,\ \e_{\lfloor n\theta_0\rfloor +1},\ldots,\e_n\sim \tilde{F},\quad F\neq\tilde{F}\]
($F,\tilde F$ unknown).
Let $\hat\e_j$ denote an estimator for the innovation $\e_j$, $j\in\zet$, to be defined below. 
We consider a Kolmogorov-Smirnov type test statistic based on the stochastic process
\begin{eqnarray}\label{hatTn}
\hat{T}_n(s,t)=\sqrt{n} \frac{\sum_{k=1}^{\lfloor ns\rfloor}w_{nk}}{n}\frac{\sum_{\ell=\lfloor ns\rfloor+1}^nw_{n\ell}}{n}\left(\hat{F}_{\ns}(t)-\hat{F}^*_{n-\lfloor ns\rfloor}(t)\right),\quad s\in [0,1],t\in\er,
\end{eqnarray}
where the sequential empirical processes are defined as
\begin{eqnarray*}
\hat{F}_{\lfloor ns\rfloor}(t)&=&\sum_{j=1}^{\lfloor ns\rfloor}\frac{w_{nj}}{\sum_{k=1}^{\lfloor ns\rfloor}w_{nk}}I\{\hat{\e}_j\leq t\}\\
\hat{F}^*_{n-\lfloor ns\rfloor}(t)&=&\sum_{j=\lfloor ns\rfloor+1}^n\frac{w_{nj}}{\sum_{k=\lfloor ns\rfloor+1}^nw_{nk}}I\{\hat{\e}_j\leq t\}.
\end{eqnarray*}
The weights are chosen as $w_{nj}=w_n(X_{j-1})$ with continuous weight function $w_n:\er\to[0,1]$ such that for some sequences $a_n\to-\infty$, $b_n\to\infty$, 
\begin{eqnarray}\label{w-ass}
w_n(x)= \begin{cases} 1, & x\in[a_n+\kappa,b_n-\kappa]\\ 0, &x\notin [a_n,b_n]\end{cases}
\end{eqnarray}
for some fixed $\kappa>0$ independent of $n$. 
The weights are included in the definition of the sequential empirical processes to avoid problems of kernel estimation in areas where only few data are available, 
compare to M³ller, Schick \& Wefelmeyer (2009) and Dette, Pardo-Fernßndez \& Van Keilegom (2009). 
Further let the residuals be defined as 
\[\he_j=\frac{X_j-\hat{m}(X_{j-1})}{\hat{\sigma}(X_{j-1})},\]
for kernel regression and variance estimators
\begin{eqnarray}\label{nw-m}
\hat{m}(x)&=&\frac{\sum_{i=1}^nK\left(\frac{x-X_{i-1}}{c_n}\right)X_i}{\sum_{i=1}^nK\left(\frac{x-X_{i-1}}{c_n}\right)}\\
\hat{\sigma}^2(x)&=&\frac{\sum_{i=1}^nK\left(\frac{x-X_{i-1}}{c_n}\right)(X_i-\hat m(x))^2}{\sum_{i=1}^nK\left(\frac{x-X_{i-1}}{c_n}\right)}
\label{nw-sigma}
\end{eqnarray}
and $\hat{\sigma}(x)=(\hat{\sigma}^2(x))^{1/2}$. Here $K$ denotes a kernel function and $c_n$ a positive sequence of bandwidths. For the ease of representation we use the same bandwidth $c_n$ to estimate $m$ and $\sigma$, though in practice it may be advisable to choose different bandwidths. The asymptotic results presented in the paper remain valid when two different bandwidths according to the assumptions (C) and (C'), respectively, in the next sections are chosen. 

We list model assumptions as well as assumptions on the estimators in the next two sections.

\section{Asymptotic results under the null hypothesis}\label{sec-asymp}
\def\theequation{3.\arabic{equation}}
\setcounter{equation}{0}

Throughout this section we make use of the following assumptions.

\begin{description}
\item [(K)] The kernel $K$ is a three times differentiable density with compact support $[-C,C]$ and $\sup_{u\in[-C,C]}|K^{(\mu)}(u)|\leq \bK <\infty$, $\mu=0,1,2,3$. Moreover $K(C)=K(-C)=K'(C)=K'(-C)=0$ and $\int K(u)udu=0$.

\item[(C)] The sequence of bandwidths $c_n$ fulfills 
$$nc_n^4(\log n)^\eta\to 0,\quad \frac{(\log n)^\eta}{nc_n^{2+\sqrt{3}}}\to 0\quad \mbox{for all } \eta>0.$$

\textit{Remark:  As can be seen from the proof the first bandwidth condition can be replaced by  
$nc_n^4(q_nq_n^\sigma)^8(q_n^f)^2=O(nc_n^4(\log n)^{8r_q+8r_s+2r_f})=o(1)$, where $q_n,q_n^\sigma,q_n^f,r_q,r_s,r_f$ are defined in assumptions (X) and (M) below. The second bandwidth condition is equivalent to the existence of some $\delta>0$ such that
\begin{eqnarray}\label{delta}
\frac{(\log n)^\eta}{nc_n^{3+2\delta}}\to 0,\quad \frac{(\log n)^\eta}{nc_n^{1+\frac{1}{\delta}}}\to 0
\end{eqnarray}
for all $\eta>0$. The first condition in (\ref{delta}) is typical in the context of empirical processes of nonparametrically estimated residuals, compare Dette, Pardo-Fernßndez \& Van Keilegom (2009) or Neumeyer \& Van Keilegom (2009), while the $\log$-factor stems from the boundary truncation via the weight function. The second condition in (\ref{delta}) arises at the very end of the proof of Lemma \ref{lemma9.1} in appendix B due to a $\delta$-dependent covering number.
The constant $\delta$ is also used in Lemma \ref{lemma3.4} in appendix B. }

\item [(I)] For the interval $I_n=[a_n,b_n]$ there exists some $r_I<\infty$ such that $(b_n-a_n)=O((\log n)^{r_I})$. Moreover $\left(\int _{-\infty}^{a_n+\kappa}f_{X_0}(x)dx+\int_{b_n-\kappa}^{\infty}f_{X_0}(x)dx\right)=o((\log n)^{-1})$.

\item [(W)] The weight function $w_n:\er\to[0,1]$ fulfills (\ref{w-ass}) and is three times differentiable such that  $\sup_{n\in\en}\sup_{x\in\er}|w_n^{(\mu)}(x)|<\infty$ for $\mu=1,2,3$.

\item [(F)] The innovations $\e_j$, $j\in\mathbb{Z}$, are identically distributed with distribution function $F$. Their density $f$ is continuously differentiable and $\sup_{t\in\er}|f(t)t|<\infty$  as well as \linebreak $\sup_{t\in\er}|f'(t)t^2|<\infty$.

\textit{Remark: Due to the continuity of the density $f$ and the derivative $f'$ it follows that also $\sup_{t\in\er}f(t)<\infty$, $\sup_{t\in\er}|f'(t)|<\infty$ and $\sup_{t\in\er}|f'(t)t|<\infty$.}

\item [(E)] There exists some $b> 1+\sqrt 3$  such that $E\left[|X_0|^{2b}\right]<\infty$ and $E\left[|\e_1|^{2b}\right]<\infty$.

\item [(X)] The observations $X_j$, $j\in\mathbb{Z}$, are identically distributed and the process $(X_j)_{j\in\zet}$ is $\alpha$-mixing with exponentially fast decaying mixing-coefficient $\alpha(n)$. \\
Their density $f_{X_0}$ is bounded and four times differentiable with bounded derivatives.
The density is also bounded away from zero on compact intervals and there exists some $r_f<\infty$ such that $q_n^f=(\inf_{x\in I_n}f_{X_0}(x))^{-1}=O((\log n)^{r_f})$.

\textit{Remark:  Assumptions (F) and (X) imply strong stationarity of the process $(X_j)_{j\in\zet}$. }

\item[(Z)]
It holds that
 $$\sup_{x\in J_n}\left(\left( |m(x)|+|\sigma(x)|\right)^{2k}f_{X_{0}}(x)\right)=O(1)$$ 
 and there exists some $1\leq j^*<\infty$ such that 
 $$\sup_{x,x'\in J_n}\left(\left(|m(x)|+|\sigma(x)|\right)^k\left(|m(x')|+|\sigma(x')|\right)^kf_{X_{0},X_{j-1}}(x,x')\right)=O(1)$$ is valid for all $j>j^*+1$,
 for $k=1,2$, $n\to\infty$ with  $J_n=[a_n-(C+c_n^{-\frac 12}n^{-\frac 12}(\log n)^{\frac 12})c_n\ ,\ b_n+(C+c_n^{-\frac 12}n^{-\frac 12}(\log n)^{\frac 12})c_n]$.

\item [(M)] The regression function $m$ and the scale function  $\sigma$ are four times differentiable and there exist some $r_q,r_s<\infty$ and $q_n$, $q_n^{\sigma}$ with $q_n=O((\log n)^{r_q})$, $q_n^{\sigma}=O((\log n)^{r_s})$, $(q_n)^{-1}=O(1)$, $(q_n^{\sigma})^{-1}=O(1)$ such that $\sup_{x\in [a_n-Cc_n,b_n+Cc_n]}|m^{(\mu)}(x)|=O(q_n)$,  \linebreak $\sup_{x\in [a_n-Cc_n,b_n+Cc_n]}|\sigma^{(\mu)}(x)|=O(q_n)$, $\mu=0,1,2,3,4$ and $(\inf_{x\in I_n}|\sigma(x)|)^{-1}=O(q_n^{\sigma})$.
\end{description}

An example for which the assumptions are fulfilled is the AR(1) model $X_j=0.5X_{j-1}+\e_j$ with standard normally distributed innovations $\e_j$, $j\in\zet$. Then the observations $X_j$, $j\in\zet$, are identically $\cn(0,\frac{4}{3})$ distributed and with $I_n=[-(\frac 83 \log((\log n)^2))^{1/2}-\kappa,  (\frac 83 \log((\log n)^2))^{1/2}+\kappa]$, a weight function that fulfills (W), a kernel function that fulfills (K) and  a bandwidth that fulfills (C) all assumptions are fulfilled. To this end note that exponential $\alpha$-mixing holds for stationary models with $\lim_{|x|\to\infty} (|m(x)|+|\sigma(x)|E[|\e_j|^\tau]^{\frac 1{\tau}})/|x|<1$  for some $\tau\geq 1$ with $E[|\e_j|^\tau]<\infty$; see Doukhan (1994).

In the first theorem we state a stochastic expansion of the residual based sequential empirical process. The proof is given in appendix A. 

\begin{theo}\label{theo1} Under model (AR) with assumptions (K), (C), (I), (W), (F), (E), (X), (Z), and (M) we have that under the null hypothesis $H_0$ of no change point 
\beq
&&\frac{1}{n}\sjns w_{nj} \left(I\left\{\hat{\e}_j\leq t\right\}-F(t)\right)\\
&=&\frac 1n\sjns\left( I\left\{\e_j\leq t\right\}-F(t)\right)+\frac{[ns]}{n}f(t)\frac 1n\sjn\e_j+\frac{[ns]}{n}f(t) t\frac 1{2n}\sjn(\e_j^2-1)+\op
\eeq
uniformly with respect to $s\in[0,1]$ and $t\in\er$.
\end{theo}

\begin{rem}\rm
The theorem complements results by M³ller, Schick \& Wefelmeyer (2009) and Dette, Pardo-Fernßndez \& Van Keilegom (2009). In both articles only non-sequential processes are considered (i.\,e.\ the case $s=1$). While M³ller, Schick \& Wefelmeyer (2009) consider a homoscedastic version of model (AR) ($\sigma\equiv const$, see also section \ref{section-hom}), Dette, Pardo-Fernßndez \& Van Keilegom (2009) consider a heteroscedastic autoregression/regression model and a result similar to Theorem \ref{theo1} (for $s=1$) can be derived from their proofs. The sequential process ($s\in [0,1]$) though requires much more involved methods of proof that also result in slightly more complicated assumptions. 
$\blacksquare$
\end{rem}

From the stochastic expansion weak convergence of the sequential residual process can be derived. The proof of Corollary \ref{cor1} is given in appendix A.

\begin{cor}\label{cor1} Under the assumptions of Theorem \ref{theo1} under the null hypothesis $H_0$ of no change point  the process 
\[\sqrt n \left(\frac{\sum_{k=1}^{\lfloor ns\rfloor}w_{nk}}n\left(\hat{F}_{\lfloor ns\rfloor}(t)-F(t)\right)\right),\quad s\in[0,1],t\in\er,\]
converges weakly to a centered Gaussian process $(\ka_F(s,t))_{s\in[0,1],t\in\er}$
with \rm
\beq
\text{Cov}(\ka_F(s_1,t_1),\ka_F(s_2,t_2))&=&s_1\wedge s_2\left(F(t_1\wedge t_2)-F(t_1)F(t_2)\right)\\
&&{}+s_1s_2\biggl(f(t_1)\left(E[\e_1I\{\e_1\leq t_2\}]+t_1E[(\e_1^2-1)I\{\e_1\leq t_2\}]\right)\\
&&\hspace{1,2cm}+f(t_2)\left(E[\e_1I\{\e_1\leq t_1\}]+t_2E[(\e_1^2-1)I\{\e_1\leq t_1\}]\right)\\
&&\hspace{1,2cm}+f(t_1)f(t_2)\left(1+(t_1+t_2)E[\e_1^3]+t_1t_2(E[\e_1^4]-1)\right)\biggr).
\eeq
\end{cor}

\begin{rem}\rm
From Theorem \ref{theo1} and Corollary \ref{cor1} it can be seen that the nonparametric estimation of the autoregression and conditional variance function vastly influences the asymptotic behaviour of the process. The asymptotic distribution of the partial sum processes decicively changes when based on residuals compared to the corresponding processes built from iid innovations. This is different from simpler situations in specific parametric time series models, see Bai (1994) and Krei▀ (1991), among others, but corresponds to situations in parametric as well as nonparametric regression models, see e.\,g.\ Koul (2002) and Neumeyer \& Van Keilegom (2009). Note however that neither the chosen kernel function nor the bandwidth have any influence on the asymptotic distribution. $\blacksquare$
\end{rem}

The stochastic expansion given in Theorem \ref{theo1} can be used to derive the asymptotic distribution of the change point test. First we state weak convergence of the process defined in (\ref{hatTn}). To this end in the following let $(\ge(s,z))_{s\in[0,1],z\in[0,1]}$ denote a completely tucked Brownian sheet, i.\,e.\ a centered Gaussian process with covariance structure
\[\text{Cov}(\ge(s_1,z_1),\ge(s_2,z_2))=(s_1\wedge s_2-s_1s_2)(z_1\wedge z_2-z_1z_2).\] 

\begin{theo}\label{theo2} Under model (AR) with the assumptions (K), (C), (I), (W), (F), (E), (X), (Z),  and (M) under the null hypothesis $H_0$ of no change point there exist  Gaussian processes $(\ge_n(s,F(t)))_{s\in[0,1],t\in\er}$, $n\in\en$,  with the same distribution as $(\ge(s,F(t)))_{s\in[0,1],t\in\er}$ such that
\beq
\sup_{s\in[0,1],t\in\er}\left|\hat T_n(s,t)-\ge_n(s,F(t))\right|&=&o_P(1).
\eeq
\end{theo}

The proof is again given in appendix A as well as the proof of the next corollary in which we state the asymptotic distribution of the change point test.

\begin{cor}\label{cor2} Under the assumptions of Theorem \ref{theo2}  under the null hypothesis $H_0$ of no change point the 
Kolmogorov-Smirnov type test statistic $\sup_{s\in[0,1],t\in\er}|\hat T_n(s,t)|$ converges in distribution to $\sup_{s\in[0,1],z\in[0,1]}|\ge(s,z)|$.
\end{cor}

\begin{rem}\rm
From Corollary \ref{cor2} it follows that the test is asymptotically distribution-free although the stochastic expansion given in Theorem \ref{theo1} still depends on the innovation distribution in a complicated way. This remarkable feature in the context of procedures based on nonparametric residual empirical processes was already observed by Neumeyer \& Van Keilegom (2009) in the context of independent observations.
The critical values for the test are tabled in Picard (1985). $\blacksquare$
\end{rem}

\section{Asymptotic results under fixed alternatives}\label{sec-alt}
\def\theequation{4.\arabic{equation}}
\setcounter{equation}{0}

The assumptions (K) and (M) as well as the following assumptions are used to proof consistency of the test under fixed alternatives.

\begin{description}
\item[(C')] The sequence of bandwidths $c_n$ fulfills 
$$nc_n^5\to 0,\quad \frac{(\log n)^\eta}{nc_n^{2+\sqrt{3}}}\to 0 \quad \mbox{for all } \eta>0.$$
\item [(I')] For the interval $I_n=[a_n,b_n]$ there exists some $r_I<\infty$  such that $(b_n-a_n)=O((\log n)^{r_I})$. 

\item [(W')] The weight function $w_n:\er\to[0,1]$ is continuous and fulfills (\ref{w-ass}).

\item [(F')]  Let $\e_j$ have distribution function $F_{\e_j}$ and density $f_{\e_j}$, $j\in\mathbb{Z}$. Let $\frac 1n\sjn\sup_{t\in\er}f_{\e_j}(t)=O(1)$  as well as $\frac 1n\sjn \sup_{t\in\er}|f_{\e_j}(t)t|=O(1)$ for $n\to\infty$.

\textit{Remark: Under the alternative $H_1$ the assumption is fulfilled when $\sup_{t\in\er} f(t)<\infty$, $\sup_{t\in\er} \tilde f(t)<\infty$, $\sup_{t\in\er} |f(t)t|<\infty$, and $\sup_{t\in\er} |\tilde f(t)t|<\infty$, where $f$ and $\tilde f$ are densities corresponding to $F$ and $\tilde F$, respectively. }

\item [(E')] It holds that $\frac 1n\sum_{i=1}^nE\left[|X_i|^{2b}\right]=O(1)$ for some $b> 1+\sqrt 3$, $n\to\infty$.

\item [(X')] The observation process $(X_j)_{j\in\zet}$ is $\alpha$-mixing with mixing-coefficient $\alpha(n)=O(n^{-\beta})$ for some \[\beta>\max\left(2\frac{(3+\sqrt 3)b+2+\sqrt 3}{(1+\sqrt 3)b-2(2+\sqrt3)}, 7\right). \]
The observation densities $f_{X_i}$  are four times differentiable and fulfill \linebreak $\sup_{x\in\er}n^{-1}\sum_{i=1}^n|f_{X_{i-1}}^{(\mu)}(x)|=O(1)$, $\mu=0,1,2,3,4$.
Moreover there exists some $r_f<\infty$ such that $q_n^f=(\inf_{x\in I_n}\frac 1n\sum_{i=1}^nf_{X_{i-1}}(x))^{-1}=O((\log n)^{r_f})$.

\item[(Z')] For all $m_n\leq n$ with $m_n^{-1}=o(1)$ it holds that 
\\ $\sup_{x\in J_n}\left(\left(|m(x)|+|\sigma(x)|\right)\ \frac 1{m_n}\max_{0\leq S\leq n-m_n}\sum_{i=S+1}^{S+m_n}f_{X_{i-1}}(x)\right)=O(1)$,\\ $\sup_{x\in J_n}\left(\left( |m(x)|+|\sigma(x)|\right)^{2k}\ \frac 1{m_n}\max_{0\leq S\leq n-m_n}\sum_{i=S+1}^{S+m_n} \left(1+E[\e_i^4]\right)^{k-1}f_{X_{i-1}}(x)\right)=O(1)$,\\\\
 and there exists some $1\leq j^*<\infty$ such that \[\sup_{x,x'\in J_n}\!\!\left(\!\!\left(|m(x)|+|\sigma(x)|\right)^k\left(|m(x')|+|\sigma(x')|\right)^k\!\frac 1{m_n^2}\max_{0\leq S\leq n-m_n}\sum_{\substack{i,j=S+1\\ |i-j|>j^*}}^{S+m_n}f_{X_{i-1},X_{j-1}}(x,x')\!\!\right)=O(1)\] and 
 \[ \sup_{x\in J_n}\left(\left(|m(x)|+|\sigma(x)|\right)^{2k}\max_{j^*+1\leq i\leq n-j^*}\sum_{j=i-j^*}^{i+j^*}\left(1+E[\e_j^4]\right)^{k-1}f_{X_{j-1}}(x)\right)=O(1)\]
 for $k=1,2$, $n\to\infty$ with  $J_n=[a_n-(C+c_n^{-\frac 12}n^{-\frac 12}(\log n)^{\frac 12})c_n\ ,\ b_n+(C+c_n^{-\frac 12}n^{-\frac 12}(\log n)^{\frac 12})c_n]$.

\textit{Remark: It suffices when the assumption is valid for $m_n$ defined in (\ref{m_n}) in the proof.\\
Under the alternative $H_1$ the arithmetic mean of all $f_{X_{i-1}}$ converges to the weighted sum of the observation density  before the change point and the long range observation density after the change point with weights $\theta_0$ and $1-\theta_0$, so (Z') is fulfilled if (Z)  and (Z) with the long range observation density instead of $f_{X_0}$ are fulfilled and the last part of (Z') holds.}

\end{description}

\begin{rem}\rm If the observations and the innovations are identically distributed it holds that  the second and third part of (X') are equivalent to the second and third part of (X) and (Z') is equivalent to (Z). The other assumptions are not equivalent, even if the the innovations are identically distributed. In detail it holds that assumption (I') is weaker than (I), (F') is weaker than (F), (E') is weaker than (E), the first part of (X') is weaker than the first part of (X), as well as (C') is weaker than (C). $\blacksquare$
\end{rem}

\begin{rem}\rm Note that under the alternative the process $(X_j)_{j\in\zet}$ is not stationary. Thus to obtain consistency most auxiliary results in appendix B are proved without assuming stationarity. A stabilisation of the density averages as in assumptions (F'), (X') is sufficient for our results to hold. In particular we generalize some of Hansen's (2008) results in Lemma \ref{lemma3.4}. $\blacksquare$
\end{rem}

\begin{theo}\label{theo3} Under the assumptions (K), (C'), (I'), (W'), (F'), (E'), (X'), (Z'), and (M) under the fixed alternative $H_1$ with a change point in  $\lfloor n\theta_0\rfloor$, we have
\begin{eqnarray*}
&&\sup_{t\in\er}\left|\frac{\sum_{k=1}^{\lfloor n\theta_0\rfloor}w_{nk}}n\left(\hat{F}_{\lfloor n\theta_0\rfloor}(t)-F(t)\right)\right|=o_P(1)\\
&&\sup_{t\in\er}\left|\frac{\sum_{k=\lfloor n\theta_0\rfloor+1}^nw_{nk}}n\left(\hat{F}^*_{n-\lfloor n\theta_0\rfloor}(t)-\tilde{F}(t)\right)\right|=o_P(1).
\end{eqnarray*}
\end{theo}

\begin{cor}\label{cor3} Under the assumptions of Theorem \ref{theo3} the Kolmogorov-Smirnov type test based on the process $\hat T_n$ is consistent against fixed alternatives $H_1$. 
\end{cor}

The proofs of Theorem \ref{theo3} and Corollary \ref{cor3} are  given in appendix A. 

\section{The homoscedastic AR-model}\label{section-hom}
\def\theequation{5.\arabic{equation}}
\setcounter{equation}{0}

In this section we consider a homoscedastic AR-model
\begin{description}
\item [(AR1)]  
\qquad\qquad $X_j=m(X_{j-1})+\e_j$,\\
where the innovations $\e_j$, $j\in\zet$, are independent with $E[\e_j]=0$ and $E[\e_j^2]<\infty$ $\forall j$ and $\e_j$ is independent of the past $X_k$, $k\leq j-1$, $\forall j$.
\end{description}
Our aim is to test the change point hypotheses $H_0$ vs.\ $H_1$ from section \ref{sec-null}. Note that here under $H_1$ the change in the innovation distribution can result from a change in the variance. 
 The residuals are now defined as $\he_j=X_j-\hat{m}(X_{j-1})$ and the test statistic is built with these in the same way as described for the heteroscedastic case; see (\ref{hatTn}). Let assumptions ($\bar{\text{Z}}$), ($\bar{\text{M}}$) under the null hypothesis and assumption (Z'') under the alternative be formulated as (Z), (M) in  section \ref{sec-asymp} and (Z') in section \ref{sec-alt}, respectively, but replacing the variance function $\sigma$ by a constant.  Let ($\bar{\text{F}}$) be formulated as (F), but replacing conditions $\sup_{t\in\er}|f(t)t|<\infty$, $\sup_{t\in\er}|f'(t)t^2|<\infty$ by $\sup_{t\in\er}f(t)<\infty$, $\sup_{t\in\er}|f'(t)t|<\infty$.
 Let (F'') be formulated as (F'), but deleting the last condition. Let ($\bar{\text{E}}$) and (E'') be formulated as (E) and (E') respectively, but replacing $2b$ by $b$. Then the following asymptotic results are valid.

\begin{theo} Under model (AR1) with assumptions (K), (C), (I), (W), ($\bar{F}$), ($\bar{E}$), (X), ($\bar{Z}$), and ($\bar{M}$)  we have that under the null hypothesis $H_0$ of no change point 
\beq
\frac{1}{n}\sjns w_{nj} \left(I\left\{\hat{\e}_j\leq t\right\}-F(t)\right)&=&
\frac 1n\sjns\left( I\left\{\e_j\leq t\right\}-F(t)\right)+\frac{[ns]}{n}f(t)\frac 1n\sjn\e_j+\op
\eeq
uniformly with respect to $s\in[0,1]$ and $t\in\er$.
\end{theo}

\begin{cor} Under the assumptions of Theorem \ref{theo1} under the null hypothesis $H_0$ of no change point  the process 
\[\sqrt n \left(\frac{\sum_{k=1}^{\lfloor ns\rfloor}w_{nk}}n\left(\hat{F}_{\lfloor ns\rfloor}(t)-F(t)\right)\right),\quad s\in[0,1],t\in\er,\]
converges weakly to a centered Gaussian process $(\ka_F(s,t))_{s\in[0,1],t\in\er}$
with \rm
\beq
&&\text{Cov}(\ka_F(s_1,t_1),\ka_F(s_2,t_2) )\;=\; s_1\wedge s_2\left( F(t_1\wedge t_2 )-F(t_1)F(t_2) \right)\\
&&{}+s_1s_2\Big( f(t_1) E [ \e_1I\{\e_1\leq t_2\} ]+f(t_2) E[\e_1I\{\e_1\leq t_1\}]+f(t_1)f(t_2)\Var(\e_1)\Big).
\eeq
\end{cor}

\begin{theo} Under model (AR1) with assumptions (K), (C'), (I'), (W'), (F''), (E''), (X'), (Z''), and ($\bar M$) the Kolmogorov-Smirnov type test based on the process $\hat T_n$ is consistent against fixed alternatives $H_1$. 
\end{theo}

The proofs are analogous to the proofs of the results in sections \ref{sec-null} and \ref{sec-alt}, but easier due to the simpler structure of the model and the process. They are omitted for the sake of brevity.

\section{Small sample performance}
\def\theequation{6.\arabic{equation}}
\setcounter{equation}{0}

\subsection{Simulations}

{\bf The heteroscedastic model.}
To examine the performance of the test on small samples we considered AR(1) models and ARCH(1) models.\\
For the AR(1) case we considered the models
\[X_j=0.5\cdot X_{j-1}+\e_j,\qquad \e_1,\ldots,\e_{\lfloor \frac n2\rfloor}\sim \cn(0,1),\ \e_{\lfloor \frac n2\rfloor+1},\ldots,\e_n\sim\tilde{F}_1\ (\text{respectively}\ \tilde{F}_2),\]
where  $\tilde{F}_1$ is the distribution function of a random variable that is $\cn(-2\zeta,1)$ distributed with probability $0.5$ and $\cn(2\zeta,1)$ distributed with probability $0.5$ and $\tilde{F}_2$  is the distribution function of a random variable that is $\cn(0,(1-\zeta)^2)$ distributed with probability $0.5$ and $\cn(0,2-(1-\zeta)^2)$ distributed with probability $0.5$, for different values of $\zeta$. 
Though the data are generated by a homoscedastic model we assume for the data analysis validity of the heteroscedastic model (AR).

\begin{figure}[h!]\begin{center}
\begin{minipage}[t]{0.47\textwidth}
\includegraphics[width=\textwidth]{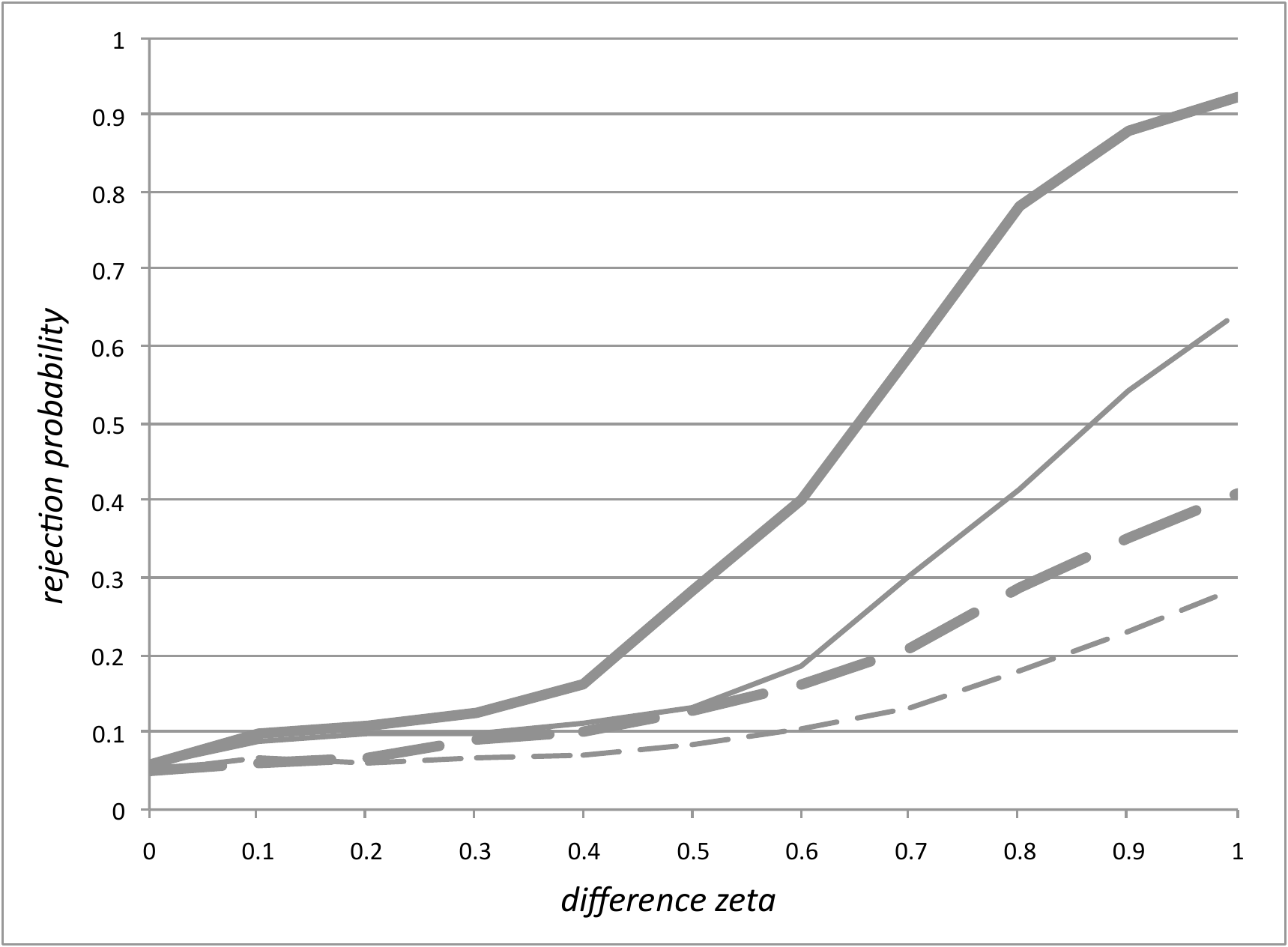}
\end{minipage}
\quad
\begin{minipage}[t]{0.47\textwidth}
\includegraphics[width=\textwidth]{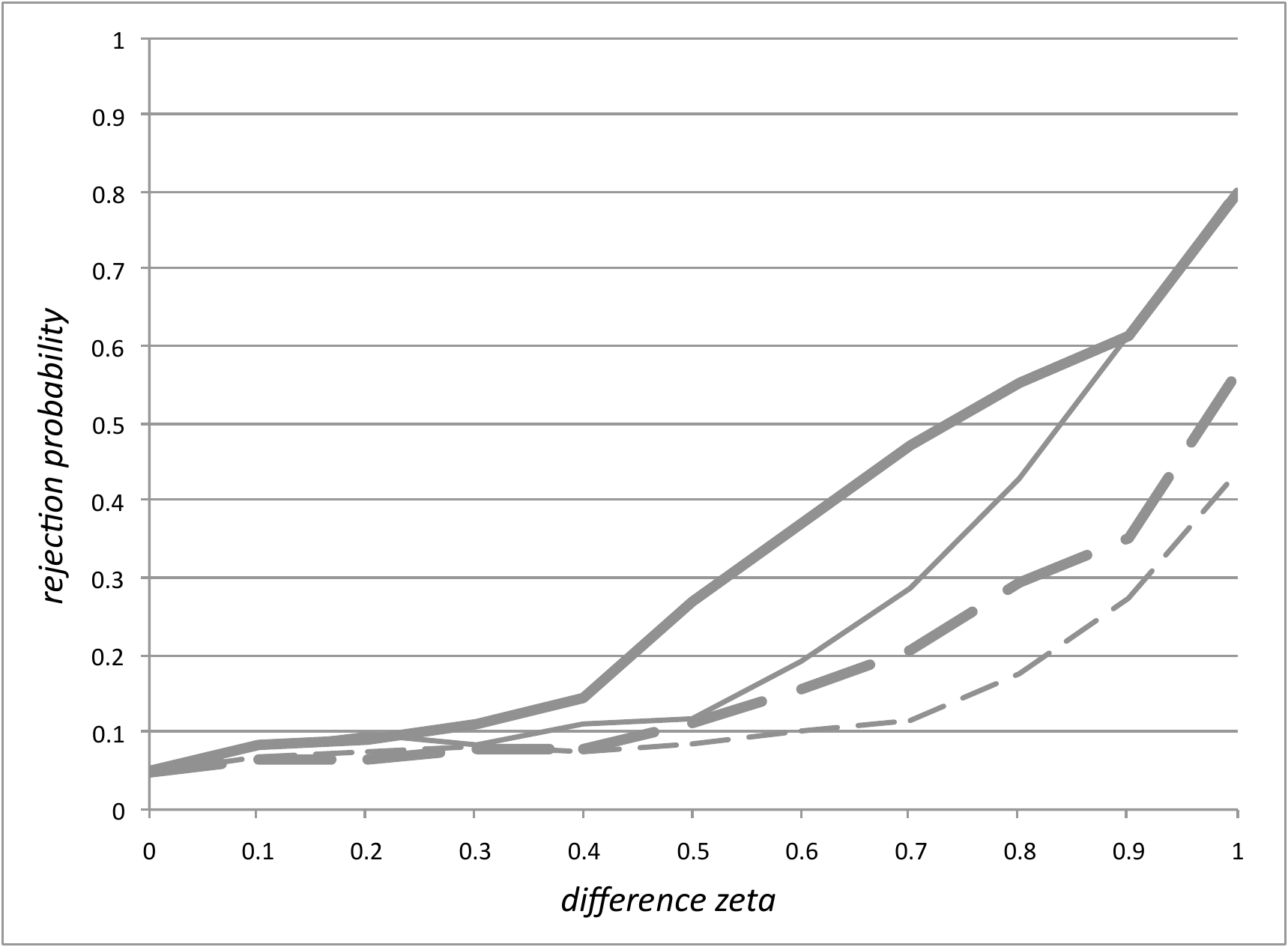}
\end{minipage}
\caption{\sl Rejection probabilities obtained from AR(1) (left) and ARCH(1) (right) models for $n=100$ (dashed curves) and $n=200$ (solid curves). The thick curves represent the results for the model with $\tilde{F}_1$, the thin curves the results for the the model with $\tilde{F}_2$.}\label{arch11b}\label{ar11b}
\end{center}\end{figure}

In Table \ref{ar11} the rejection probabilities for 500 repetitions, level 5\% and sample sizes $n\in\{100,200\}$ are shown. They are also shown in the left panel of Figure \ref{ar11b}.
It can be seen that the level is approximated well and the power increases for increasing parameter $\zeta$ as well as for increasing sample size $n$. 

\begin{table}[h!]\begin{center}
\begin{tabular}{| l || c | c | c | c | c | c | c | c |}
\hline
 $\quad\%$&$\ \zeta=0\ $&$\zeta=0.1$&$\zeta=0.2$&$\zeta=0.3$&$\zeta=0.4$&$\zeta=0.6$&$\zeta=0.8$&$\zeta=1$\\
\hline\hline
$\tilde{F}_1$, $n=100$&$5$&$6.2$&$7$&$9.2$&$10.2$&$16.4$&$29$&$41$\\
\hline
$\tilde{F}_1$, $n=200$&$5.8$&$9.8$&$10.6$&$12.4$&$16.2$&$40.2$&$78.2$&$92.4$\\
\hline\hline
&&&&&&&& $\zeta=0.99$\\
\hline\hline
$\tilde{F}_2$, $n=100$&$5$&$6.8$&$6.4$&$7$&$7.2$&$10.8$&$18$&$28.2$\\
\hline
$\tilde{F}_2$, $n=200$&$5.8$&$8.8$&$9.8$&$9.6$&$11.2$&$18.4$&$41.6$&$63.4$\\
\hline
\end{tabular}
\caption{\sl Rejection probabilities obtained from AR(1) models}\label{ar11}
\end{center}\end{table}

The same kind of change points was examined for ARCH(1) models
\[X_j=\sqrt{0.75+0.25X_{j-1}^2}\cdot\e_j,\qquad \e_1,\ldots,\e_{\lfloor \frac n2\rfloor}\sim \cn(0,1),\ \e_{\lfloor \frac n2\rfloor+1},\ldots,\e_n\sim\tilde{F}_1\ (\text{respectively}\ \tilde{F}_2)\]
with rejection probabilities for 500 repetitions and level 5\% as displayed in Table \ref{arch11} and in the right panel of Figure \ref{arch11b}.

\begin{table}[h!]\begin{center}
\begin{tabular}{| l || c | c | c | c | c | c | c | c |}
\hline
 $\quad\%$&$\ \zeta=0\ $&$\zeta=0.1$&$\zeta=0.2$&$\zeta=0.3$&$\zeta=0.4$&$\zeta=0.6$&$\zeta=0.8$&$\zeta=1$\\
\hline\hline
$\tilde{F}_1$, $n=100$&$4.8$&$6.4$&$6.6$&$7.8$&$8$&$15.6$&$29.6$&$56.8$\\
\hline
$\tilde{F}_1$, $n=200$&$5$&$8.4$&$9.2$&$11.2$&$14.6$&$37.2$&$55.4$&$80$\\
\hline\hline
&&&&&&&& $\zeta=0.99$\\
\hline\hline
$\tilde{F}_2$, $n=100$&$4.8$&$6.8$&$7.6$&$8$&$7.4$&$10.2$&$17.6$&$42.2$\\
\hline
$\tilde{F}_2$, $n=200$&$5$&$8.2$&$10$&$8.6$&$11.4$&$19.4$&$43.2$&$78.8$\\
\hline
\end{tabular}
\caption{\sl Rejection probabilities obtained from ARCH(1) models}\label{arch11}
\end{center}\end{table}

\begin{figure}[h!]\begin{center}
\begin{minipage}[t]{0.47\textwidth}
\includegraphics[width=\textwidth]{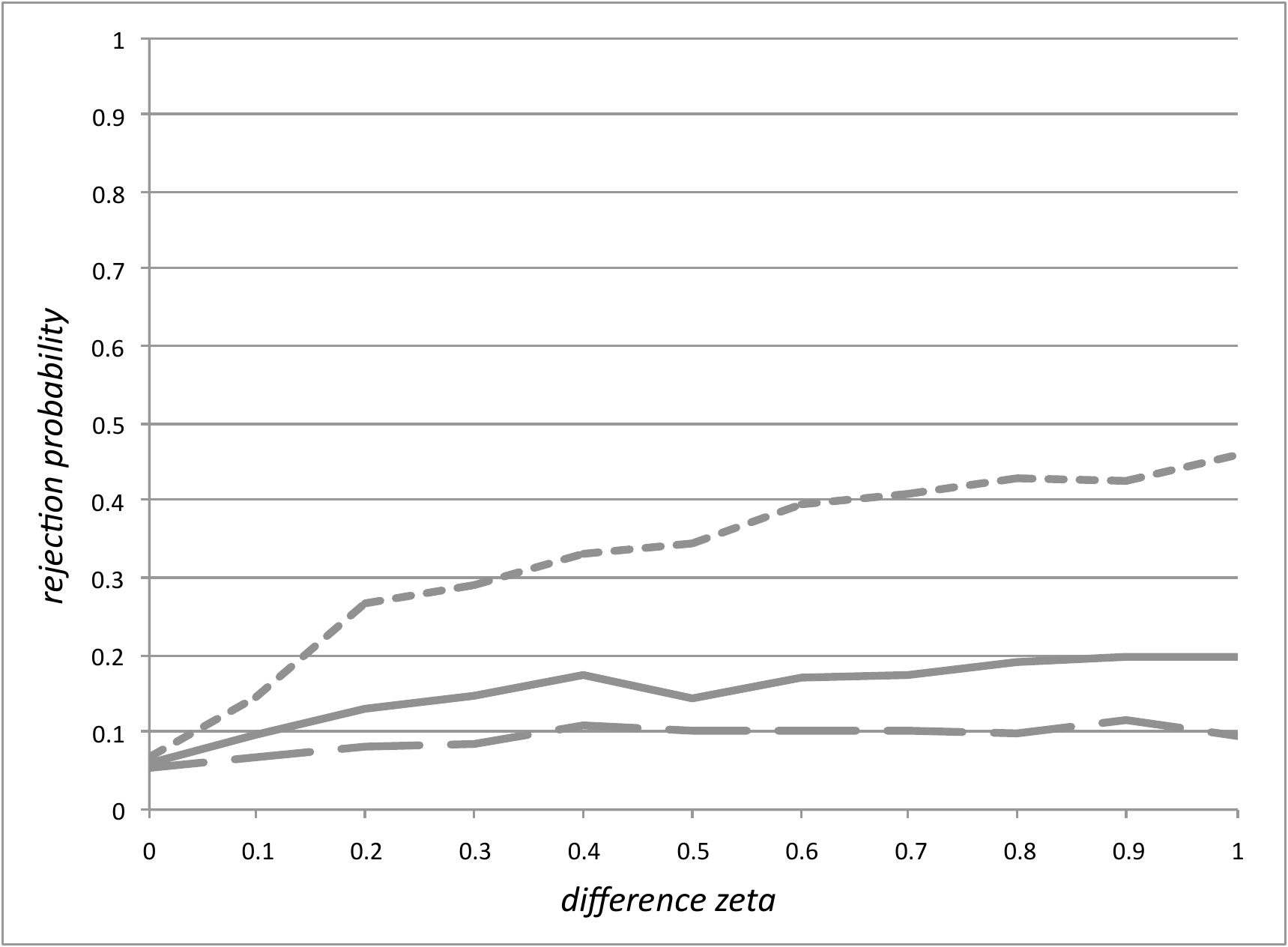}
\caption{\sl Rejection probabilities obtained from ARCH(1) models with $\mathcal{S}t(3+10\zeta)$  distributed innovations for $n=100$ (dashed curve), $n=200$ (solid curve) and $n=500$ (dotted curve).}\label{arch12b}
\end{minipage}
\quad
\begin{minipage}[t]{0.47\textwidth}
\includegraphics[width=\textwidth]{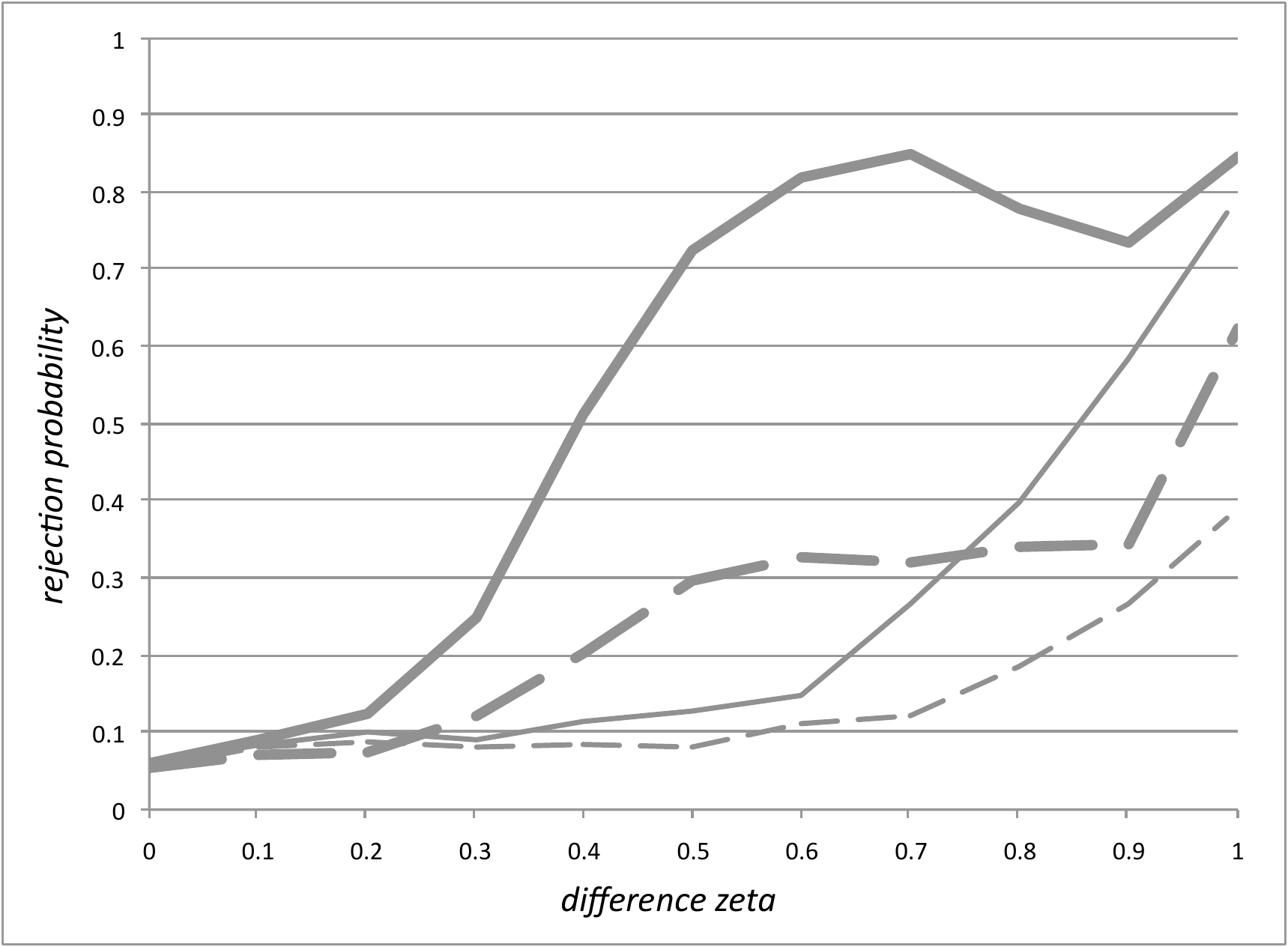}
\caption{\sl Rejection probabilities obtained from ARCH(1) models for $n=100$ (dashed curve) and $n=200$ (solid curve). The thick curves represent the results for the model with $\tilde{F}_3$, the thin curves the results for the the model with $\tilde{F}_4$.}\label{arch13b}
\end{minipage}
\end{center}\end{figure}

We also considered innovations with Student-t distribution with three degrees of freedom. The Student-t distribution has heavier tails than the normal distribution and is therefore more appropriate for modeling financial data. Due to the fact that Var$(\e_j)$ has to be one for all $j$, the Student-t distribution was standardized. 
We considered the ARCH(1) models
\[X_j=\sqrt{0.75+0.25X_{j-1}^2}\cdot\e_j,\qquad \e_1,\ldots,\e_{\lfloor \frac n2\rfloor}\sim \mathcal{S}t(3),\ \e_{\lfloor \frac n2\rfloor+1},\ldots,\e_n\sim\mathcal{S}t(3+10\zeta)\]
for different values of $\zeta$. 
The rejection probabilities for 500 repetitions, level 5\% and sample sizes $n\in\{100,200,500\}$ are shown in Table \ref{arch12} and in Figure \ref{arch12b}.

\begin{table}[h!]\begin{center}
\begin{tabular}{| l || c | c | c | c | c | c | c | c |}
\hline
 $\quad\%$&$\ \zeta=0\ $&$\zeta=0.1$&$\zeta=0.2$&$\zeta=0.3$&$\zeta=0.4$&$\zeta=0.6$&$\zeta=0.8$&$\zeta=1$\\
\hline\hline
$n=100$&$5.4$&$6.8$&$8.2$&$8.4$&$10.6$&$10$&$9.6$&$9.4$\\
\hline
$n=200$&$6$&$9.6$&$12.8$&$14.6$&$17.2$&$17$&$18.8$&$19.8$\\
\hline
$n=500$&$6.8$&$14.6$&$26.6$&$28.8$&$33$&$39.4$&$42.8$&$46$\\
\hline
\end{tabular}
\caption{\sl Rejection probabilities obtained from ARCH(1) models with $\mathcal{S}t(3+10\zeta)$  distributed innovations}\label{arch12}
\end{center}\end{table}

The asymptotic level is approximated reasonably well and the power increases with increasing $\zeta$ as well as with increasing $n$. 
Here the increase with $\zeta$ for small $n$ is not as pronounced as for the models considered before, because the difference between the distribution before and after the change point is for $\zeta=0.5$ just slightly different to that for $\zeta=1$ because the Student-t distribution converges to the standard normal distribution.
\\
We also examined the following ARCH(1) models:
\[X_j=\sqrt{0.75+0.25X_{j-1}^2}\cdot\e_j,\qquad \e_1,\ldots,\e_{\lfloor \frac n2\rfloor}\sim \mathcal{S}t(3),\ \e_{\lfloor \frac n2\rfloor+1},\ldots,\e_n\sim\tilde{F}_3\ (\text{respectively}\ \tilde{F}_4), \]
where  $\tilde{F}_3$ is the distribution function of a random variable that is $\left(\mathcal{S}t(3)-2\zeta\right)$ distributed with probability $0.5$ and $\left(\mathcal{S}t(3)+2\zeta\right)$ distributed with probability $0.5$ and $\tilde{F}_4$  is the distribution function of a random variable that is $(1-\zeta)\cdot\mathcal{S}t(3)$ distributed with probability $0.5$ and $\sqrt{2-(1-\zeta)^2}\cdot\mathcal{S}t(3)$ distributed with probability $0.5$, for different values of $\zeta$.\\
The rejection probabilities for 500 repetitions and level 5\% are shown in Table \ref{arch13} and Figure \ref{arch13b}.

\begin{table}[h!]\begin{center}
\begin{tabular}{| l || c | c | c | c | c | c | c | c |}
\hline
 $\quad\%$&$\ \zeta=0\ $&$\zeta=0.1$&$\zeta=0.2$&$\zeta=0.3$&$\zeta=0.4$&$\zeta=0.6$&$\zeta=0.8$&$\zeta=1$\\
\hline\hline
$\tilde{F}_3$, $n=100$&$5.4$&$7$&$7.4$&$12.2$&$20.2$&$32.8$&$34$&$62.4$\\
\hline
$\tilde{F}_3$, $n=200$&$6$&$9.2$&$12.4$&$25$&$51.2$&$82$&$78$&$84.6$\\
\hline\hline
&&&&&&&& $\zeta=0.99$\\
\hline\hline
$\tilde{F}_4$, $n=100$&$5.4$&$8$&$8.8$&$8.2$&$8.4$&$11.2$&$18.6$&$37.8$\\
\hline
$\tilde{F}_4$, $n=200$&$6$&$8.6$&$10.2$&$9.2$&$11.6$&$15$&$39.8$&$77.4$\\
\hline
\end{tabular}
\caption{\sl Rejection probabilities obtained from ARCH(1) models}\label{arch13}
\end{center}\end{table}

The models with Student-t distributed innovations with three degrees of freedom do not fulfill the moment assumptions, because moments greater than or equal to $3$ do not exist, but the simulations show that the test works on them just the same.

\begin{figure}[h!]\begin{center}
\begin{minipage}[t]{0.47\textwidth}
\includegraphics[width=\textwidth]{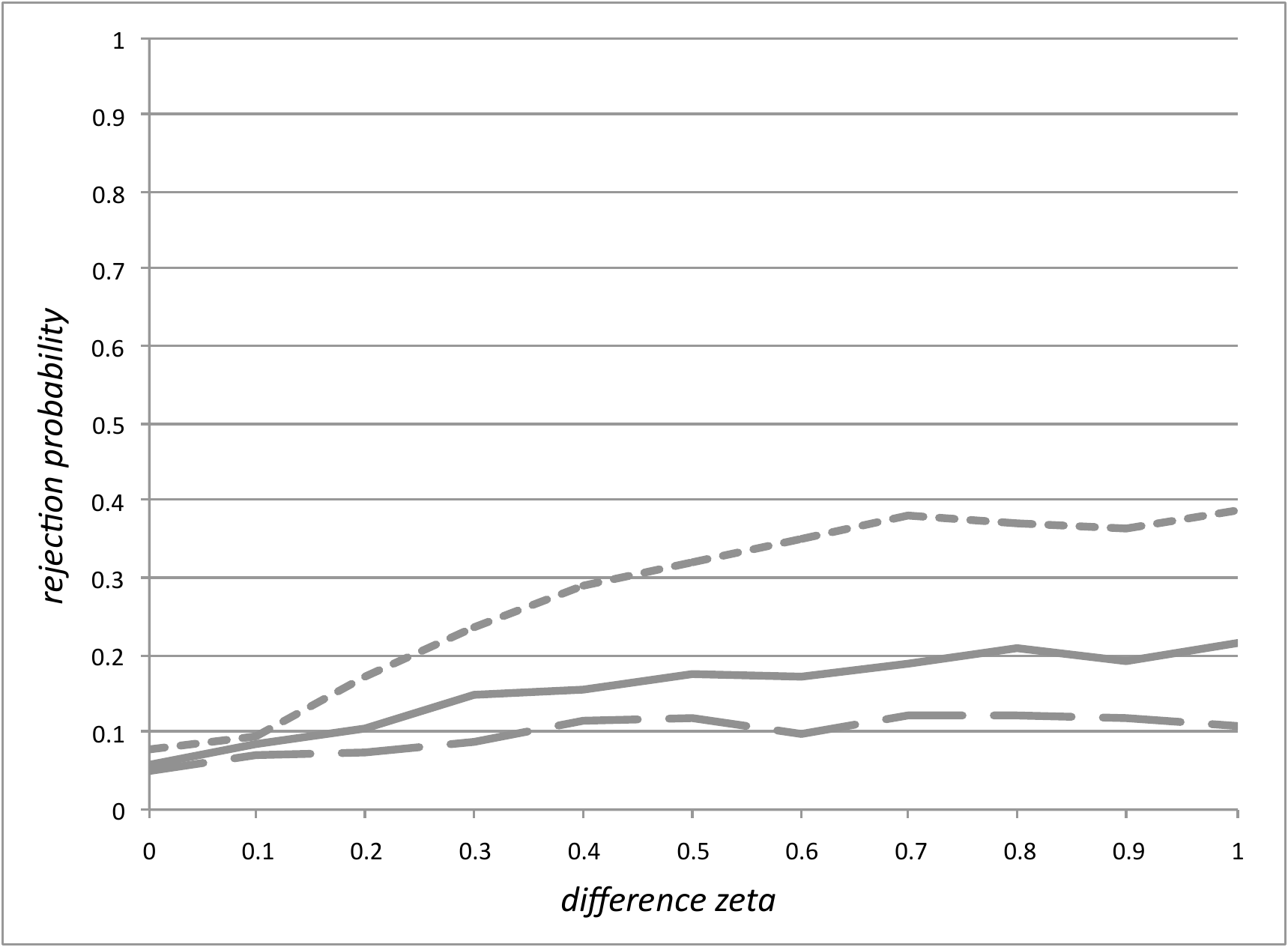}
\end{minipage}
\quad
\begin{minipage}[t]{0.47\textwidth}
\includegraphics[width=\textwidth]{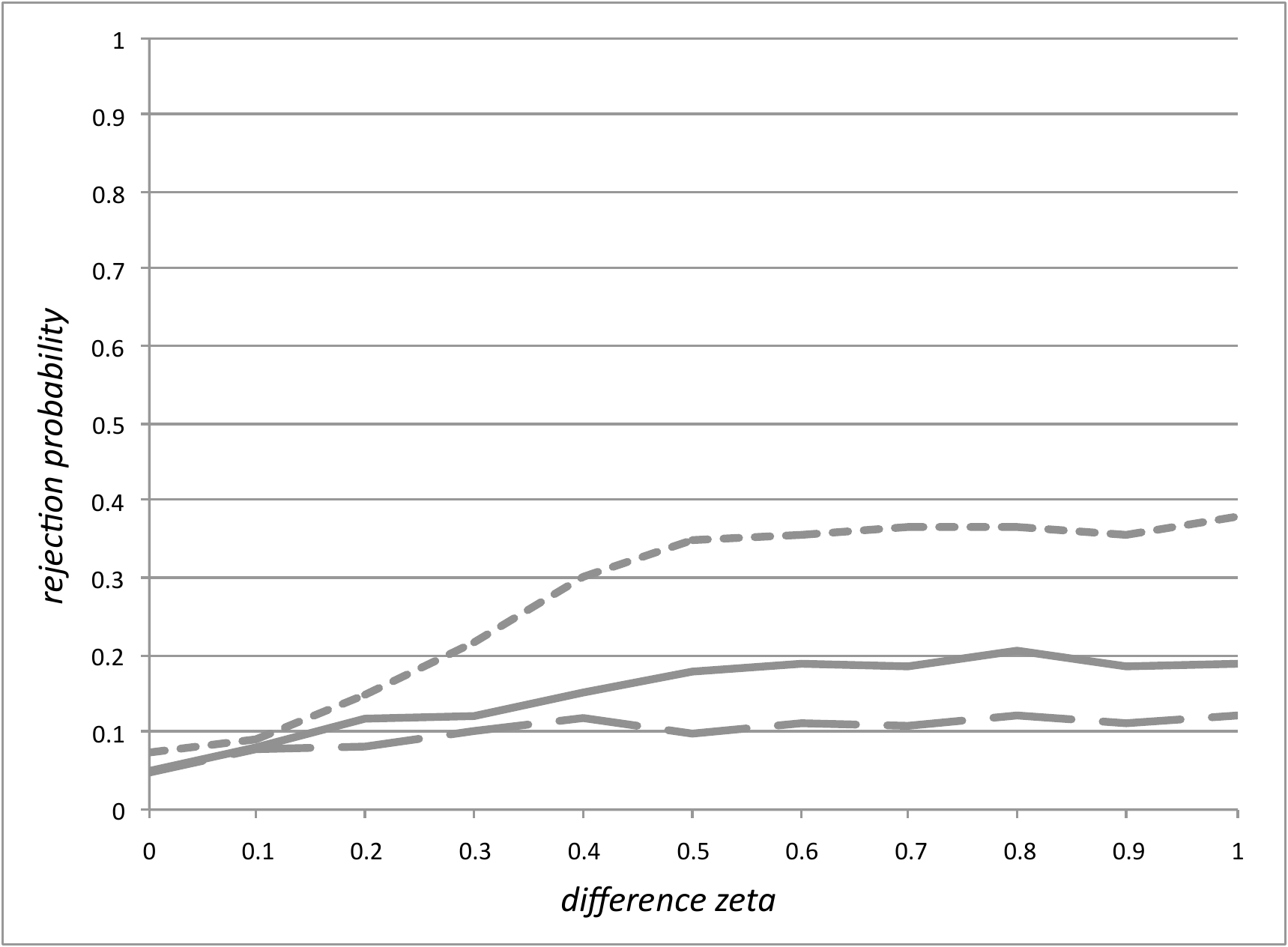}
\end{minipage}
\caption{\sl Rejection probabilities obtained from AR(1) (left) and ARCH(1) (right) models with skew-normal distributed innovations for $n=100$ (dashed curve), $n=200$ (solid curve) and $n=500$ (dotted curve).}\label{ar1arch1b}
\end{center}\end{figure}

Finally, we considered the  skew-normal distribution as innovation distribution. 
Let $\tilde F_5$ denote the skew-normal distribution with location parameter $$-\sqrt{\frac{2\pi\left(\left(10\zeta\right)^2+\left(10\zeta\right)^4\right)}{\pi^2+\left(2\pi^2-2\pi\right)\cdot\left(10\zeta\right)^2+\left(\pi^2-2\pi\right)\cdot\left(10\zeta\right)^4}},$$ scale parameter
$(\pi(1+(10\zeta)^2)^{1/2}/(\pi+(\pi-2)(10\zeta)^2)^{1/2}$ and shape parameter $10\zeta$.
We considered the AR(1) and ARCH(1) models
\[X_j=0.5\cdot X_{j-1}+\e_j,\qquad \e_1,\ldots,\e_{\lfloor \frac n2\rfloor}\sim \cn(0,1),\ \e_{\lfloor \frac n2\rfloor+1},\ldots,\e_n\sim\tilde{F}_5\]
and
\[X_j=\sqrt{0.75+0.25X_{j-1}^2}\cdot\e_j,\qquad \e_1,\ldots,\e_{\lfloor \frac n2\rfloor}\sim \cn(0,1),\ \e_{\lfloor \frac n2\rfloor+1},\ldots,\e_n\sim\tilde{F}_5\]
for different values of $\zeta$. The parameters in the skew-normal distribution were chosen like this to guarantee $E[\e_j]=0$ and Var$(\e_j)=1$ for all $j$.
The rejection probabilities for 500 repetitions and level 5\% are shown in Table \ref{ar1arch1} and Figure \ref{ar1arch1b}.

\begin{table}[h!]\begin{center}
\begin{tabular}{| l || c | c | c | c | c | c | c | c |}
\hline
 $\quad\%$&$\zeta=0$&$\zeta=0.1$&$\zeta=0.2$&$\zeta=0.3$&$\zeta=0.4$&$\zeta=0.6$&$\zeta=0.8$&$\zeta=1$\\
\hline\hline
\small AR(1), $n=100$&$5$&$7$&$7.2$&$8.6$&$11.4$&$9.8$&$12.2$&$10.8$\\
\hline
\small AR(1), $n=200$&$5.8$&$8.4$&$10.2$&$14.8$&$15.2$&$17.2$&$20.8$&$21.6$\\
\hline
\small AR(1), $n=500$&$7.8$&$9.2$&$17.2$&$23.4$&$28.8$&$34.8$&$36.8$&$38.8$\\
\hline\hline
\small ARCH(1), $n=100$&$4.8$&$7.6$&$8$&$10$&$11.8$&$11$&$12$&$12.2$\\
\hline
\small ARCH(1), $n=200$&$5$&$7.8$&$11.8$&$12$&$15.2$&$18.8$&$20.6$&$18.6$\\
\hline
\small ARCH(1), $n=500$&$7.4$&$9$&$14.8$&$21.8$&$30$&$35.4$&$36.6$&$38$\\
\hline
\end{tabular}
\caption{\sl Rejection probabilities obtained from AR(1) and ARCH(1) models with skew-normal distributed innovations}\label{ar1arch1}
\end{center}\end{table}

\bigskip

\begin{figure}[h!]\begin{center}
\includegraphics[width=0.47\textwidth]{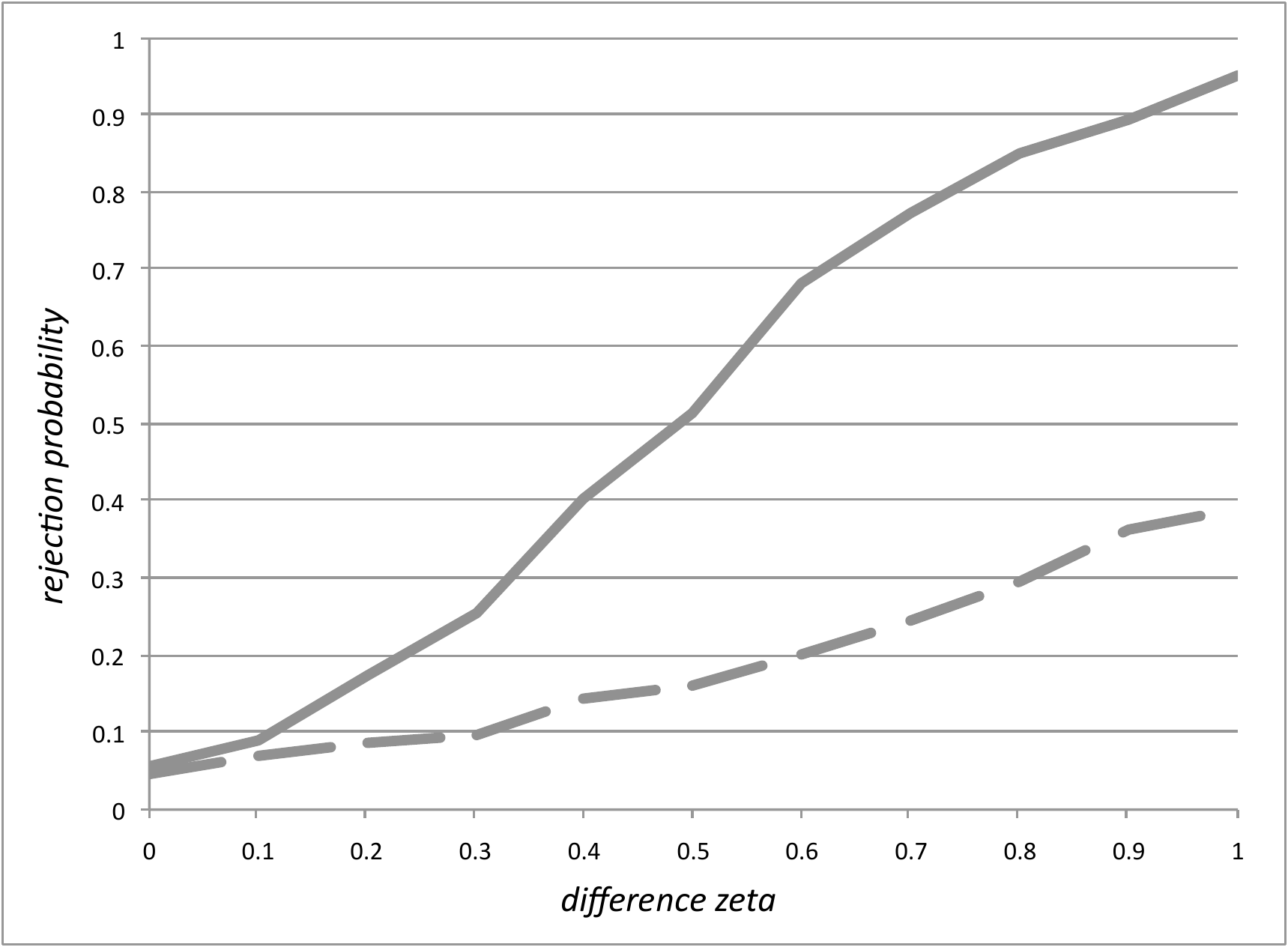}
\caption{\sl Rejection probabilities obtained from homoscedastic AR(1) models with change in variance for $n=100$ (dashed curve) and $n=200$ (solid curve).}\label{ar13b}
\end{center}\end{figure}

{\bf The homoscedastic model.} 
For the homoscedastic model $X_j=m(X_{j-1})+\e_j$ as considered in section \ref{section-hom} only Var$(\e_j)<\infty$ $\forall j$ is assumed so that we
can simulate a change in the variance. To this end we generated data from the  AR(1) model
\[X_j=0.5\cdot X_{j-1}+\e_j,\qquad \e_1,\ldots,\e_{\lfloor \frac n2\rfloor}\sim \cn(0,0.5^2),\ \e_{\lfloor \frac n2\rfloor+1},\ldots,\e_n\sim\cn(0,(0.5+\zeta)^2)\]
for different values of $\zeta$.\\
The rejection probabilities for 500 repetitions and level 5\% are shown in Table \ref{ar13} and in Figure \ref{ar13b}.

\begin{table}[h!]\begin{center}
\begin{tabular}{| l || c | c | c | c | c | c | c | c |}
\hline
 $\quad\%$&$\ \zeta=0\ $&$\zeta=0.1$&$\zeta=0.2$&$\zeta=0.3$&$\zeta=0.4$&$\zeta=0.6$&$\zeta=0.8$&$\zeta=1$\\
\hline\hline
$n=100$&$4.6$&$6.8$&$8.6$&$9.8$&$14.4$&$20$&$29.4$&$39$\\
\hline
$n=200$&$5.6$&$9$&$17.2$&$25.2$&$40.2$&$68.2$&$85.2$&$95.2$\\
\hline
\end{tabular}
\caption{\sl Rejection probabilities obtained from homoscedastic AR(1) models with change in variance}\label{ar13}
\end{center}\end{table}

It can be seen that the theoretical results are supported by the simulations.

\bigskip

\begin{figure}[h!]\begin{center}
\begin{minipage}[t]{0.47\textwidth}
\includegraphics[width=\textwidth]{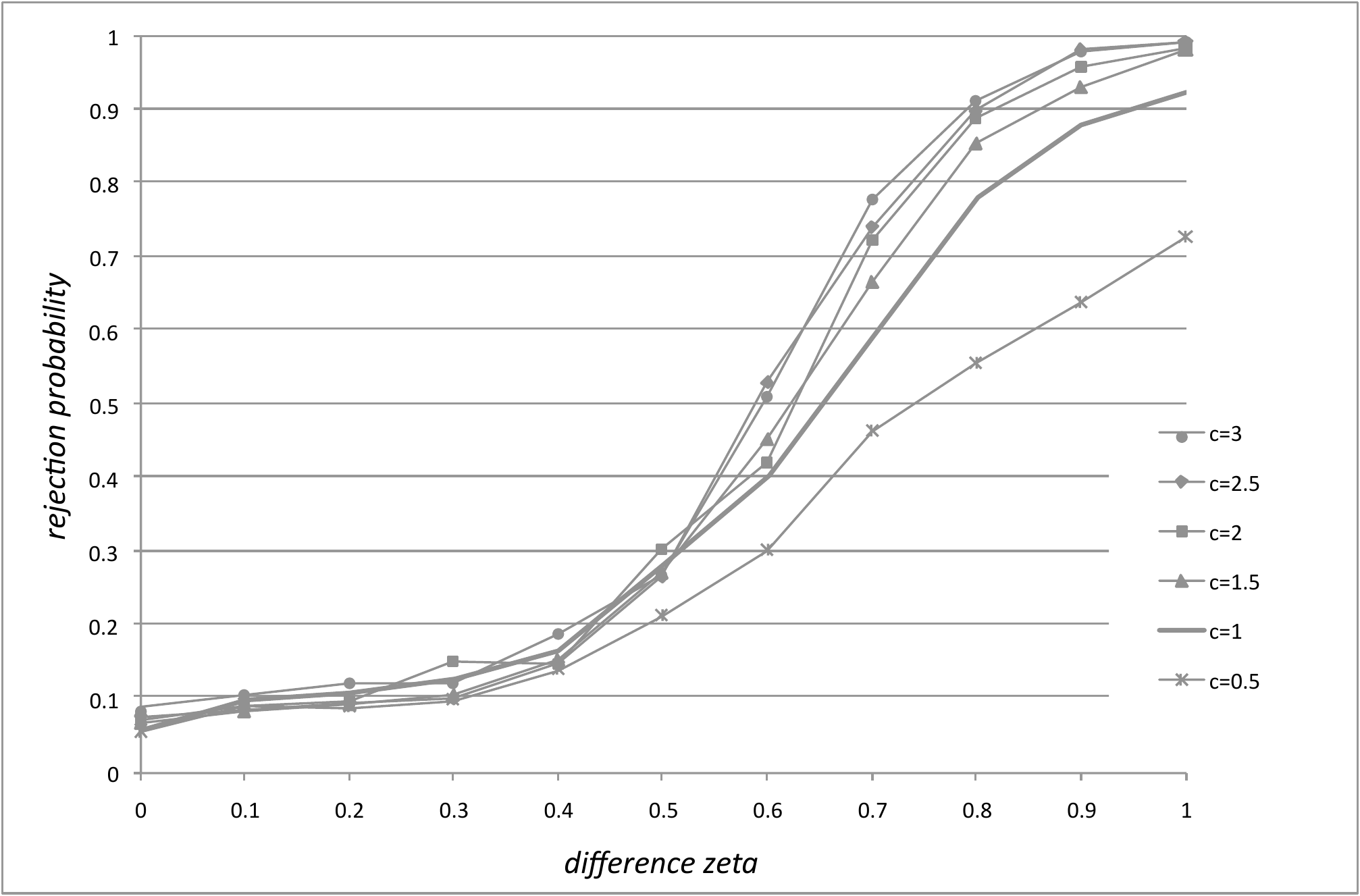}
\end{minipage}
\quad
\begin{minipage}[t]{0.47\textwidth}
\includegraphics[width=\textwidth]{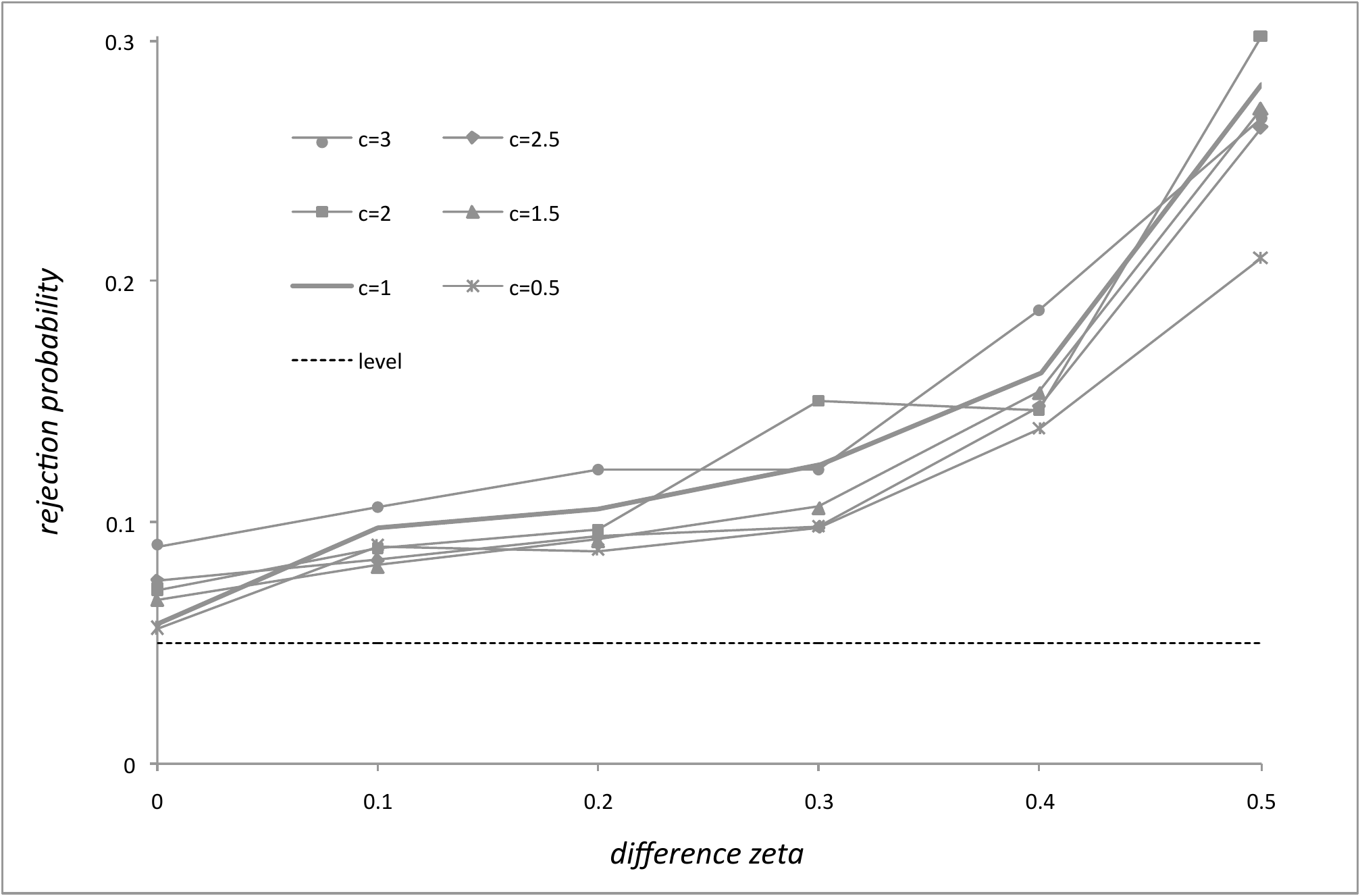}
\end{minipage}
\caption{\sl Rejection probabilities obtained form AR(1) models with $\tilde{F}_1$ for different sizes of bandwidth and $n=200$.}\label{bandwidthb}
\end{center}\end{figure}

{\bf Simulation setting.}
For each simulation $10\cdot n$ observations $X_j$ were generated, $9.5\cdot n$ with distribution before and $0.5\cdot n$ with distribution after the change point. For the test the last $n$ observations were used. This was done to ensure that the process is in balance.\\
The empirical processes were built without the weight function $w_n$, which means that $I_n$ was chosen as the real line. This is contrary to the assumptions. Nevertheless the simulations support our theoretical results very well, so it can be assumed that the weight function is necessary for the theory but the test can be used regardless.\\
The Nadaraya-Watson estimators $\hat{m}$ and $\hat{\sigma}$ were calculated with Gaussian kernel and bandwidth $c_n=n^{-\frac 14}$. This is also not compatible with all assumptions, e.\,g.\ the support of the kernel is not compact. However this has negligible effect on the simulations because the Gaussian kernel decreases exponentially fast at the tails. The choice of bandwidth is not compatible to the assumption as well because it does not converge faster than $n^{-\frac 14}$. A compatible choice would be $c_n=n^{-\frac 14}(\log n)^{-r}$ for some adequate $0<r<\infty$, but for the small sample sizes that were used the logarithm would be too strong in comparison to $n^{-\frac 14}$, so we omitted it.\\
To study the influence of the size of bandwidth we simulated the first AR(1) model (with $\tilde{F}_1$) with $c_n=c\cdot n^{-\frac 14}$ for different values of $c\in\er_{>0}$. The results are shown in Figure \ref{bandwidthb}. It can be seen that the rejection probability increases with $c$, especially for $\zeta\geq 0.5$, but also the rejection probability under the null hypothesis increases with $c$.

\subsection{Real data applications}

\begin{figure}[h!]\begin{center}
\begin{minipage}[t]{\textwidth}
\includegraphics[width=\textwidth]{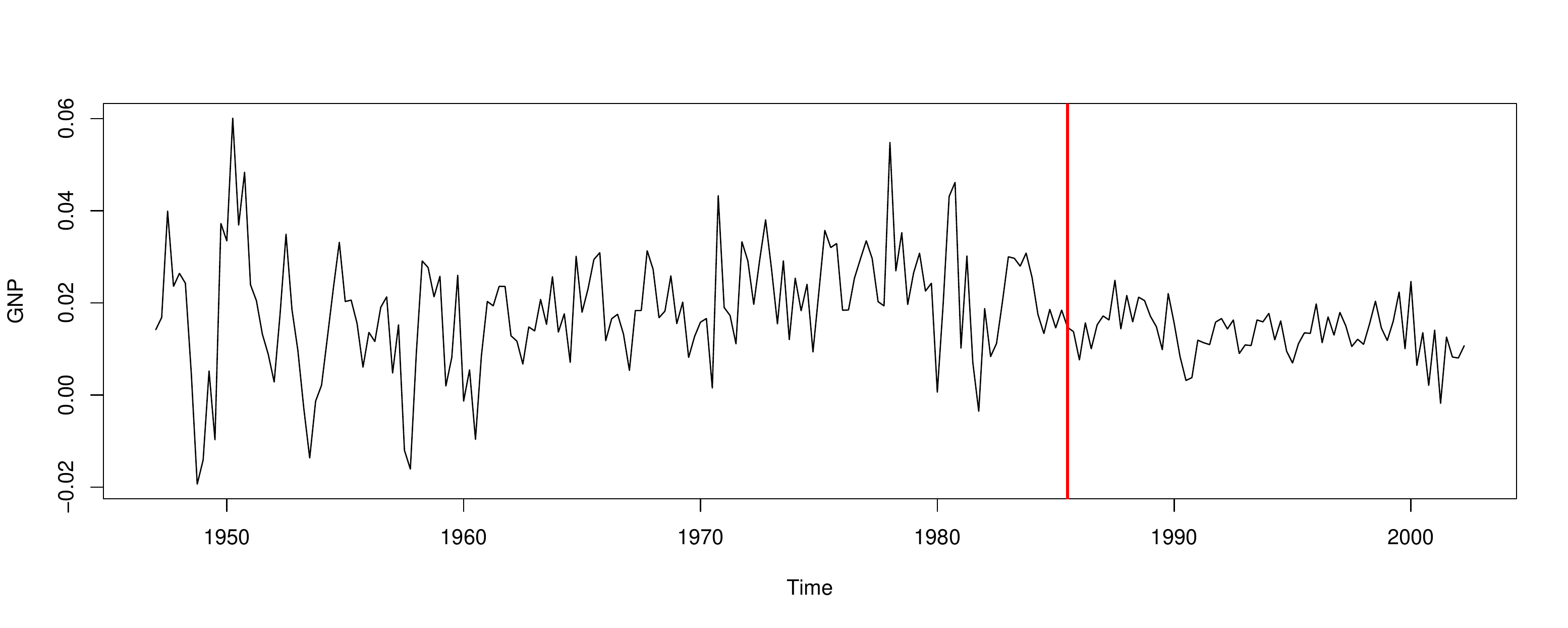}
\caption{\sl U.S. GNP quarterly growth rate}\label{gnp_2002}
\end{minipage}

\begin{minipage}[t]{\textwidth}
\includegraphics[width=\textwidth]{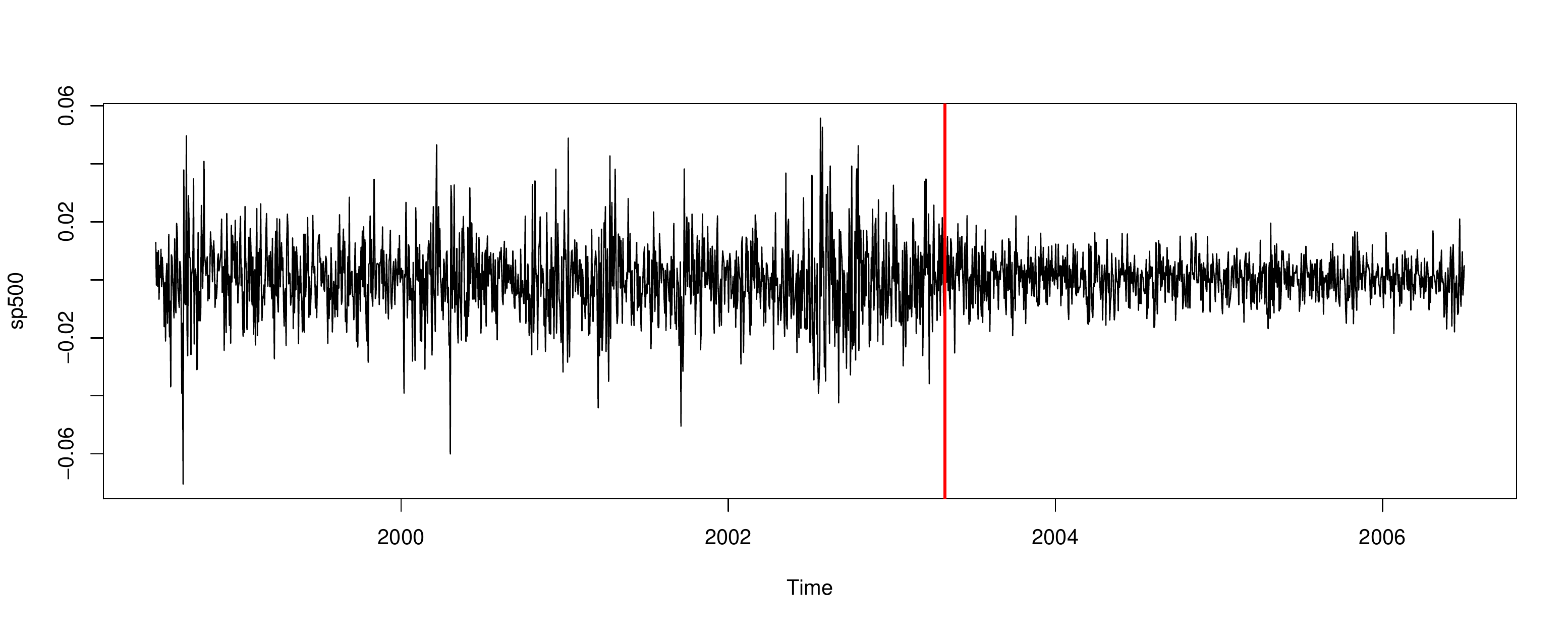}
\caption{\sl S\&P 500 daily log-return}\label{sp500}
\end{minipage}
\end{center}\end{figure}

We also applied our new test to real datasets. Firstly we examined the quarterly GNP (Gross National Product) of the USA in billions of dollars from 1947(1) to 2002(3). The data have been seasonally adjusted. We looked at the difference of the logarithm of the GNP, which is naturally interpreted as the growth rate of GNP. Figure \ref{gnp_2002} shows that there might be a structural break in the data and indeed our testing procedure rejects the null hypothesis of no change point with p-value smaller than 0.001. The vertical line marks the point $\ns$ at which the test process $\hat{T}_n(s,t)$ is maximal. We used the test statistic for homoscedastic cases, next to the one for heteroscedastic cases, for these data, because the plot suggests that there might be some change in the variance. Both tests delivered the same value of the test statistic which is 1.392 (approximately).
\\
The same data were examined in Shao \& Zhang (2010) with some kind of CUSUM test that is based on an self-normalization method. They tested for a possible change in the marginal variance, 75\% quantile and 25\% quantile  of the observations, and the test for a change in the 75\% quantile rejected the null hypothesis of no change point with p-value smaller than 0.001. The tests for a change in the marginal variance and 25\% quantile did not reject the null hypothesis of no change point. The p-values for these were greater than 0.1.
\\
Shumway \& Stoffer (2006) also examined these data and used stationary time series models, such as AR(1) and MA(2), to fit them. Both model fits pass their diagnostic checking tests, but our results, as well as the results of Shao \& Zhang (2010), indicate that the data might not be a stable process but contain a change point.

Another dataset that we examined is the daily log-return of the S\&P 500 index, a world known stock index that is quoted at the New York stock exchange, from July 1st 1998 to June 30th 2006. Figure \ref{sp500} shows that there might be a structural break in these data as well, which is confirmed by our testing procedure with p-value smaller than 0.001. Again the vertical line marks the point $\ns$ at which the test process $\hat{T}_n(s,t)$ is maximal. Like in the GNP example we used the test statistics for both cases (hetero- and homoscedastic) and both delivered nearly the same value of the test statistic which is 1.578 for the heteroscedastic and 1.575 for the homoscedastic case (approximately). 
\\
We examined these data although it is known that for financial data often higher moments do not exist, because our simulation study with Student-t distributed innovations shows that the testing procedure works even if the moment assumption is not fulfilled.
\\
The S\&P 500 data were also examined by Kirch \& Tadjuidje Kamgaing (2012). They used a testing procedure based on cumulative sums of parametrically estimated residuals for nonlinear autoregressive models. Instead of log-returns they used transformed squared log-returns and they also reject the null hypothesis of no change point. They do not give a p-value but reject clearly at level 5\%.

\section{Concluding remarks and outlook}

In this paper we transferred classical ideas of testing for change points in samples of independent observations to testing for change points in the innovation distribution in nonparametric autoregressive models with conditional heteroscedasticity.  To this end we considered the sequential empirical process of estimated innovations, proved an asymptotic expansion and weak convergence. We showed that the classical Kolmogorov-Smirnov test for a change point is asymptotically distribution-free in the new context and is not influenced asymptotically by the nonparametric estimation of the innovations. We proved consistency of the test under fixed alternatives and demonstrated the good performance in a simulation study. The proofs are based on empirical process theory for time series data and require the development of several technical auxiliary results. In particular we prove uniform rates for kernel estimators and their derivatives under nonstationarity assumptions. 

It is the topic of a future project to apply the  theory  developed here to test for serial independence of innovations or independence of the current innovation and past observations resp.\ covariates in nonparametric time series regression models. Moreover our aim is to model $k$-dimensional joint innovation distributions in multivariate time series models.

\def\theequation{7.\arabic{equation}}
\setcounter{equation}{0}

\begin{appendix}

\section{Proofs: main results}
\def\theequation{A.\arabic{equation}}
\setcounter{equation}{0}

In this section we give the proofs for Theorems \ref{theo1}, \ref{theo2}, \ref{theo3} and Corollaries \ref{cor1}, \ref{cor2}, \ref{cor3}, whereas some auxiliary results (Lemmata \ref{lemma3.3}--\ref{lemma9.4}) are stated and proved in section B. 
In some proofs standard arguments are given in condensed form for the sake of brevity. All details can be found in Selk (2011).  

\subsection{Proofs for results under the null hypothesis}

{\bf Proof of Theorem \ref{theo1}.}
From Lemma \ref{lemma9.1} and Lemma \ref{lemma9.3} it follows that under $H_0$ uniformly with respect to $s\in[0,1]$ and $t\in\er$
\begin{eqnarray}\label{9.1-1}
\frac{1}{n}\sjns w_{nj} \left(I\left\{\hat{\e}_j\leq t\right\}-F(t)\right)
&=& \frac{1}{n}\sjns w_{nj} \left(I\left\{{\e}_j\leq t\right\}-F(t)\right)+ R_n(s,t)+\op,\nn\\
&=& \frac{1}{n}\sjns \left(I\left\{{\e}_j\leq t\right\}-F(t)\right)+ R_n(s,t)+\op,\qquad
\end{eqnarray}
where for
\begin{eqnarray}\label{Rn}
R_n(s,t) &=&\frac{1}{n}\sjns w_{nj}\left(F\left(\frac{\hat{m}-m}{\sigma}(X_{j-1})+t\frac{\hat{\sigma}}{\sigma}(X_{j-1})\right)-F(t)\right)
\end{eqnarray}
it is straightforward to show by a first order Taylor expansion applying assumption (F) as well as Lemma \ref{lemma3.4} that
\begin{eqnarray}\label{rn-entw}
R_n(s,t) &=&f(t)\frac{1}{n}\sjns w_{nj}\left(\frac{\hat{m}-m}{\sigma}(X_{j-1})+t\frac{\hat{\sigma}}{\sigma}(X_{j-1})-t\right)+\op
\end{eqnarray}
uniformly with respect to $s$ and $t$.
Now inserting the definition of $\hat m$ from (\ref{nw-m}) we have
\begin{eqnarray*}
\frac{1}{n}\sjns w_{nj}\frac{\hat{m}-m}{\sigma}(X_{j-1})&=&
\frac{1}{n}\sum_{i=1}^n \eps_i\sigma(X_{i-1})\frac{\sjns w_{nj}  K\left(\frac{X_{j-1}-X_{i-1}}{c_n}\right)\frac{1}{\sigma(X_{j-1})}}{\sum_{k=1}^nK\left(\frac{X_{j-1}-X_{k-1}}{c_n}\right)}\\\nb
&&{}+ \frac{1}{n}\sum_{i=1}^n \frac{\sjns w_{nj}  K\left(\frac{X_{j-1}-X_{i-1}}{c_n}\right)\frac{(m(X_{i-1})-m(X_{j-1}))}{\sigma(X_{j-1})}}{\sum_{k=1}^nK\left(\frac{X_{j-1}-X_{k-1}}{c_n}\right)}
\\
&=& \frac{1}{n}\sum_{i=1}^n \eps_i\frac 1n \sjns w_{nj}+\op 
\end{eqnarray*}
uniformly with respect to $s$ and $t$, which follows from Lemma \ref{lemma-10} (i)--(iii).
Now 
\begin{eqnarray}\label{cheb}
\sup_{s\in [0,1]} \left|\fnsk-\fns\right|&=&\frac 1n\sjn \left(1-w_n(X_{j-1})\right)\;=\; o_P(1)
\end{eqnarray}
can be shown by Chebyshev's inequality and we obtain
\begin{eqnarray} \label{rn-entw2}
\frac{1}{n}\sjns w_{nj}\frac{\hat{m}-m}{\sigma}(X_{j-1})&=&
\frac{1}{n}\sum_{i=1}^n \eps_i\frac{[ns]}{n}+\op .
\end{eqnarray}
By an application of
$$\frac{\hat\sigma}{\sigma}-1= \frac{\hat\sigma^2-\sigma^2}{2\sigma^2}-\frac{(\hat\sigma-\sigma)^2}{2\sigma^2}$$
and Lemma \ref{lemma3.4}, for the second term arising from (\ref{rn-entw}) we have
\begin{eqnarray*}
\frac{1}{n}\sjns w_{nj}\left(\frac{\hat{\sigma}}{\sigma}(X_{j-1})-1\right)&=&
\frac{1}{n}\sjns w_{nj}\frac{\hat{\sigma}^2-\sigma^2}{2\sigma^2}(X_{j-1})+\op.
\end{eqnarray*}
Now noting that 
\begin{eqnarray}\label{sig-dar}
\hat{\sigma}^2(x)&=&\frac{\sum_{i=1}^nK(\frac{x-X_{i-1}}{c_n})X_i^2}{\sum_{l=1}^nK(\frac{x-X_{l-1}}{c_n})}-(\hat{m}(x))^2
\end{eqnarray}
 similarly to the derivation of (\ref{rn-entw2}) with results similar to those in Lemma \ref{lemma-10} one obtains
\begin{eqnarray}
\frac{1}{n}\sjns w_{nj}\left(\frac{\hat{\sigma}}{\sigma}(X_{j-1})-1\right)&=&
\frac 1{2n}\sjn(\e_j^2-1)\frac{[ns]}{n}+\op.\label{rn-entw3}
\end{eqnarray}
Finally from (\ref{rn-entw}), (\ref{rn-entw2}) and (\ref{rn-entw3}) one has
$$R_n(s,t)=\frac{[ns]}{n}f(t)\frac 1n\sjn\e_j+\frac{[ns]}{n}f(t) t\frac 1{2n}\sjn(\e_j^2-1)+\op.$$
and the assertion follows from this equality and (\ref{9.1-1}).
\hfill $\Box$

\medskip

\noindent {\bf Proof of Corollary \ref{cor1}.}
By an application of Theorem \ref{theo1} it remains to show that
\begin{eqnarray}
\nn&&\sqrt n\left(\frac 1n\sjns\left( I\left\{\e_j\leq t\right\}-F(t)\right)+\frac{\lfloor ns\rfloor}{n}f(t)\frac 1n\sjn\e_j+\frac{\lfloor ns\rfloor}{n}f(t) t\frac 1{2n}\sjn(\e_j^2-1)\right)\\
&=&\sjn \bigl( Z_{nj}(s,t,f(t),f(t)t)-E\left[Z_{nj}(s,t,f(t),f(t)t)\right]\bigr)\label{cpschwachekonvergenzsequentiell2}
\end{eqnarray}
converges weakly to the process $(\ka_F(s,t))_{s\in[0,1],t\in\er}$.
Here we use the notations
\[Z_{nj}(s,t,u,v)=n^{-\frac 12}\left(I\left\{\e_j\leq t\right\}I\left\{\frac jn\leq s\right\}+\fns u\e_j+\fns\frac 12 v(\e_j^2-1)\right),\quad (s,t,u,v)\in\cf,\]
where
\[\cf=\left\{(s,t,u,v):\  s\in[0,1],t\in\er, u\in\left[0,\sup_{t\in\er}f(t)\right], v\in \left[-\sup_{t\in\er}|f(t)t|,\sup_{t\in\er}|f(t)t|\right]\right\}\]
is equipped with the semi-metric 
\[\rho((s,t,u,v),(s',t',u',v'))=|s-s'|+|F(t)-F(t')|+|u-u'|+|v-v'|\]
and is totally bounded. 
By an application of Theorem 2.11.9 in van der Vaart \& Wellner (1996) one can show weak convergence of the process 
$$\left(\sjn \bigl( Z_{nj}(s,t,u,v)-E\left[Z_{nj}(s,t,u,v)\right]\bigr)\right)_{(s,t,u,v)\in\cf}$$
to a centered Gaussian process. Details are omitted for the sake of brevity, but the arguments are similar to (but simpler than) those in the proofs in Neumeyer \& Van Keilegom (2009) (see their proof of theorem 3 and the online supporting information). 

The assertion now follows by the continuous mapping theorem applied to the projections $u=f(t)$ and $v=f(t)t$ and by a straightforward calculation of the asymptotic covariance. 
\hfill $\Box$

\medskip

\noindent {\bf Proof of Theorem \ref{theo2}.} For the process $\hat{T}_n$ defined in (\ref{hatTn}) we have by a straightforward calculation
\begin{eqnarray*}
&&\hat{T}_n(s,t)\\
&=&\sqrt n\left(\fnskm\frac 1n\sjns w_{nj}I\left\{\hat{\e}_j\leq t\right\}-\fnsk\frac 1n\sjnsm w_{nj}I\left\{\hat{\e}_j\leq t\right\}\right)\\
&=&\sqrt n\left(\frac{\sum_{k=1}^nw_{nk}}{n}\frac 1n\sjns w_{nj}\left(I\left\{\hat{\e}_j\leq t\right\}-F(t)\right)-\fnsk\frac 1n\sjn w_{nj}\left(I\left\{\hat{\e}_j\leq t\right\}-F(t)\right)\right)\\
&=&\sqrt n\left(\frac 1n\sjns w_{nj}\left(I\left\{\hat{\e}_j\leq t\right\}-F(t)\right)-\fns\frac 1n\sjn w_{nj}\left(I\left\{\hat{\e}_j\leq t\right\}-F(t)\right)\right)+o_P(1)
\end{eqnarray*}
uniformly with respect to $s$ and $t$. 
Here the last equality follows from (\ref{cheb}). 
 Now inserting the expansion given in Theorem \ref{theo1} we directly obtain 
\begin{eqnarray*}
&&\hat{T}_n(s,t)\\
&=&
\sqrt n\fns\left(1-\fns\right)\left(\frac 1{\ns}\sjns I\left\{\e_j\leq t\right\}-\frac 1{n-\ns}\sjnsm I\left\{\e_j\leq t\right\}\right)+o_P(1)
\end{eqnarray*}
uniformly with respect to $s$ and $t$. 
The assertion follows from Remark 2 in Neumeyer \& Van Keilegom (2009), which goes back to Theorem 3.1 by Cs\"org\"o, Horvßth \& Szyszkowicz (1997).
\hfill $\Box$

\medskip

\noindent {\bf Proof of Corollary \ref{cor2}.}
It directly follows from Theorem \ref{theo2} and the continuous mapping theorem that
\[\sup_{s\in[0,1],t\in\er}\left|\hat{T}_n(s,t)\right|\xrightarrow[n\to\infty]{\cd} \sup_{s\in[0,1],t\in\er}\left|\ge(s,F(t))\right| \;=\; \sup_{s\in[0,1],z\in[0,1]}\left|\ge(s,z)\right|\]
where the last equality holds by continuity of $F$. 
\hfill $\Box$

\subsection{Proofs for results under fixed alternatives}

\noindent {\bf Proof of Theorem \ref{theo3}.} Assume that $H_1$ is valid with a change point in $\lfloor n\theta_0\rfloor$. Analogously to the proof of Theorem \ref{theo1} we have from Lemma \ref{lemma9.1}  that uniformly with respect to $t\in\er$
\begin{eqnarray*}
\frac{\sum_{k=1}^{\lfloor n\theta_0\rfloor}w_{nk}}n\left(\hat{F}_{\lfloor n\theta_0\rfloor}(t)-F(t)\right)
&=&\frac{1}{n}\sum_{j=1}^{\lfloor n\theta_0\rfloor}w_{nj} (I\{\he_j\leq t\}-F(t))\\
&=& \frac{1}{n}\sum_{j=1}^{\lfloor n\theta_0\rfloor}w_{nj} \left(I\left\{{\e}_j\leq t\right\}-F(t)\right)+ R_n(\theta_0,t)+o_P(1)\\
&=& R_n(\theta_0,t)+o_P(1),
\end{eqnarray*}
 where the last equality follows from Lemma \ref{lemma9.4}. Here, for $R_n$ as defined in (\ref{Rn}) one has by the mean value theorem applying assumption (F') and Lemma \ref{lemma3.4} that
\begin{eqnarray*}
\sup_{t\in\er}|R_n(\theta_0,t)| &\leq&
\sup_{x\in I_n}\left|\frac{\hat{m}-m}{\sigma}(x)\right|\sup_{t\in\er}|f(t)|
+\sup_{x\in I_n}\left|\frac{\hat{\sigma}-\sigma}{\sigma}(x)\right|\sup_{t\in\er}|f(t)t|O_P(1)
\;=\;o_P(1).
\end{eqnarray*}
Thus the first assertion of the Theorem follows. The second assertion is shown analogously. 
\hfill $\Box$

\medskip

\noindent {\bf Proof of Corollary \ref{cor3}.} Assume that $H_1$ is valid with a change point in $\lfloor n\theta_0\rfloor$. For $\hat T_n$ defined in (\ref{hatTn}) we have 
\begin{eqnarray*}
\sup_{t\in\er}\left|\hat{T}_n(\theta_0,t)\right|&=&\sqrt n\sup_{t\in\er}\left|\frac{\sum_{k=1}^{\nt}w_{nk}}n\frac{\sum_{k=\lfloor n\theta_0\rfloor+1}^nw_{nk}}n\left(\hat{F}_{\lfloor n\theta_0\rfloor}(t)-\hat{F}^*_{n-\lfloor n\theta_0\rfloor}(t)\right)\right|\\
&\geq & \sqrt{n}\left(\sup_{t\in\er}\left|\frac{\sum_{k=1}^{\nt}w_{nk}}n\frac{\sum_{k=\lfloor n\theta_0\rfloor+1}^nw_{nk}}n\left(F(t)-\tilde{F}(t)\right)\right|-o_P(1)\right)
\end{eqnarray*}
from Theorem \ref{theo3}. Now under $H_1$ the right hand side converges to infinity in probability from which consistency follows. 
\hfill $\Box$

\section{Proofs: auxiliary results}
\def\theequation{B.\arabic{equation}}
\setcounter{equation}{0}

\subsection{Results}

For easy overview we first state the auxiliary results and then collect the proofs in the next subsection. 

\begin{lemma}\label{lemma3.3} Under the assumptions of either Theorem \ref{theo1} or Theorem \ref{theo3} we have that
\[\sup_{x\in I_n}\left|\frac 1{nc_n}\sum_{i=1}^nK^{(\nu)}\left(\frac{x-X_{i-1}}{c_n}\right)X_i^k-E\left[\frac 1{nc_n}\sum_{i=1}^nK^{(\nu)}\left(\frac{x-X_{i-1}}{c_n}\right)X_i^k\right]\right|=O_P(\be)\]
for  $\nu=0,1,2$, $k=0,1,2$, where $\be=c_n^{-\frac 12}n^{-\frac 12}(\log n)^{\frac 12}$.
\end{lemma}

\begin{lemma}\label{lemma3.4} Under the assumptions of either Theorem \ref{theo1} or Theorem \ref{theo3} we have that
\begin{itemize}
\item[\bf (i)]
$\ds\sup_{x\in I_n}\left|\frac{\hat{m}(x)-m(x)}{\sigma(x)}\right|=O_P\left(\left(c_n^{-\frac 12}n^{-\frac 12}(\log n)^{\frac 12}+c_n^2\right)q_nq_n^fq_n^{\sigma}\right)=o_P(1)$,

$\ds\sup_{x\in I_n}\left|\frac{\hat{\sigma}(x)-\sigma(x)}{\sigma(x)}\right|=O_P\left(\left(c_n^{-\frac 12}n^{-\frac 12}(\log n)^{\frac 12}+c_n^2\right)(q_nq_n^fq_n^{\sigma})^2\right)=o_P(1)$, where $q_n$, $q_n^f$, $q_n^\sigma$ are defined in (M) and (X) resp.\ (X'). 
\item[\bf (ii)]
$\ds\sup_{x\in I_n}\left|\frac{\partial}{\partial x}\left(\frac{\hat{m}(x)-m(x)}{\sigma(x)}\right)\right|=o_P(1)$,
$\ds\sup_{x\in I_n}\left|\frac{\partial}{\partial x}\left(\frac{\hat{\sigma}(x)-\sigma(x)}{\sigma(x)}\right)\right|=o_P(1)$
\item[\bf (iii)]
$\ds\sup_{x,y\in I_n,x\neq y}\frac{\left|\frac{\partial}{\partial x}\left(\frac{\hat{m}(x)-m(x)}{\sigma(x)}\right)-\frac{\partial}{\partial y}\left(\frac{\hat{m}(y)-m(y)}{\sigma(y)}\right)\right|}{|y-x|^{\delta}}=o_P(1)$,\\
$\ds\sup_{x,y\in I_n,x\neq y}\frac{\left|\frac{\partial}{\partial x}\left(\frac{\hat{\sigma}(x)-\sigma(x)}{\sigma(x)}\right)-\frac{\partial}{\partial y}\left(\frac{\hat{\sigma}(y)-\sigma(y)}{\sigma(y)}\right)\right|}{|y-x|^{\delta}}=o_P(1)$ with $\delta$ from (\ref{delta}).\\
\end{itemize}
\end{lemma}

\begin{lemma}\label{lemma-10}
Under the assumptions of Theorem \ref{theo1} we have
\begin{itemize}
\item[\bf (i)] 
$\ds \sup_{s\in [0,1]}\left|\frac 1{n^2}\sum_{i=1}^n\e_i\sjns w_{nj}\left(\frac{\frac {1}{c_n}K\left(\frac{X_{j-1}-X_{i-1}}{c_n}\right)-\frac 1{nc_n}\sum_{k=1}^nK\left(\frac{X_{j-1}-X_{k-1}}{c_n}\right)}{\frac 1{nc_n}\sum_{k=1}^nK\left(\frac{X_{j-1}-X_{k-1}}{c_n}\right)}\right)\right|=\op$
\item[\bf (ii)]
$\ds \sup_{s\in [0,1]}\left|\frac 1{n^2}\sum_{i=1}^n\e_i\sjns\frac{w_{nj}}{\sigma(X_{j-1})}\left(\frac{\frac {1}{c_n}K\left(\frac{X_{j-1}-X_{i-1}}{c_n}\right)(\sigma(X_{i-1})-\sigma(X_{j-1}))}{\frac 1{nc_n}\sum_{k=1}^nK\left(\frac{X_{j-1}-X_{k-1}}{c_n}\right)}\right)\right|=\op$
\item[\bf (iii)]
$\ds \sup_{s\in [0,1]}\left|\frac 1n\sjns\frac{w_{nj}}{\sigma(X_{j-1})}\left(\frac{\frac 1{nc_n}\sum_{i=1}^nK\left(\frac{X_{j-1}-X_{i-1}}{c_n}\right)(m(X_{i-1})-m(X_{j-1}))}{\frac 1{nc_n}\sum_{k=1}^nK\left(\frac{X_{j-1}-X_{k-1}}{c_n}\right)}\right)\right|=\op$.\\
\end{itemize}
\end{lemma}

\begin{lemma}\label{lemma9.1} Under the assumptions  of either Theorem \ref{theo1} or Theorem \ref{theo3}  we have that
\beq
\frac 1n\sjns w_{nj}\!\left(I\{\he_j\leq t\}-F_{\e_j}\!\!\left(t\frac{\hat{\sigma}}{\sigma}(X_{j-1})+\frac{\hat{m}-m}{\sigma}(X_{j-1})\right)\!-I\{\e_j\leq t\}+F_{\e_j}(t)\right)&=& o_p\left(\frac 1{\sqrt n}\right)
\eeq
uniformly with respect to $s\in[0,1]$ and $t\in\er$, where under $H_0$ all $F_{\eps_j}$ are equal to $F$.
\end{lemma}

\begin{lemma}\label{lemma9.3} Under the assumptions of Theorem \ref{theo1} we have that
\beq
\frac{1}{n}\sjns w_{nj} \left(I\left\{{\e}_j\leq t\right\}-F(t)\right)
&=&\frac{1}{n}\sjns \left(I\left\{{\e}_j\leq t\right\}-F(t)\right)
+\op
\eeq
uniformly with respect to $s\in[0,1]$ and $t\in\er$.
\end{lemma}

\begin{lemma}\label{lemma9.4} Under the assumptions of Theorem \ref{theo3} we have that
\beq
\frac{1}{n}\sum_{j=1}^{\lfloor n\theta_0\rfloor} w_{nj} \left(I\left\{{\e}_j\leq t\right\}-F(t)\right)
\;=\; o_P(1), &&
\frac{1}{n}\sum_{j=\lfloor n\theta_0\rfloor+1}^n w_{nj} \left(I\left\{{\e}_j\leq t\right\}-\tilde F(t)\right)
\;=\; o_P(1)
\eeq
uniformly with respect to $t\in\er$.
\end{lemma}

\subsection{Proofs}

\noindent {\bf Proof of Lemma \ref{lemma3.3}.} 
Let $k\in\{0,1,2\}$. 
Throughout the proof we assume that $|X_i^k|\leq n^{\frac 1b}\log n$ for all $i$ with $b$ from assumptions (E), (E'). This is possible, because
\beq
P\left(\max_{1\leq i\leq n}\left|X_i^k\right|>n^{\frac 1b}\log n\right)&\leq&\sum_{i=1}^nP\left(\left|X_i^k\right|>n^{\frac1b}\log n \right)\;\leq\;\frac 1{n(\log n)^b}\sum_{i=1}^n E\left[|X_i|^{kb}\right]\;=\; o(1)
\eeq
by assumption (E) resp.\ (E'). 

Choose points $x_j$, $j=1,\dots,M_n^*\leq M_n=(b_n-a_n)/(\be c_n)$ (for $I_n=[a_n,b_n]$ from assumption (I) resp.\ (I')) such that $I_n$ is covered by intervals $[x_j-\be c_n,x_j+\be c_n]$.
Now let $\bK_I=\bK I_{[-C,C]}$, where $K^{(\nu+1)}$ is bounded by $\bK$ on the support $[-C,C]$ of $K$ for $\nu\in\{0,1,2\}$ (assumption (K)). Then by the mean value theorem we have
\beq
&&\left|\frac 1{nc_n}\sum_{i=1}^nK^{(\nu)}\left(\frac{x-X_{i-1}}{c_n}\right)X_i^k-E\left[\frac 1{nc_n}\sum_{i=1}^nK^{(\nu)}\left(\frac{x-X_{i-1}}{c_n}\right)X_i^k\right]\right|\\
&\leq&\left|\frac 1{nc_n}\sum_{i=1}^nK^{(\nu)}\left(\frac{x_j-X_{i-1}}{c_n}\right)X_i^k-E\left[\frac 1{nc_n}\sum_{i=1}^nK^{(\nu)}\left(\frac{x_j-X_{i-1}}{c_n}\right)X_i^k\right]\right|\\
&&+\be\frac 1{nc_n}\sum_{i=1}^n \bK_I\left(\frac{x_j-X_{i-1}}{c_n}\right)\left|X_i^k\right|\ +\ \be E\left[\frac 1{nc_n}\sum_{i=1}^n \bK_I\left(\frac{x_j-X_{i-1}}{c_n}\right)\left|X_i^k\right|\right]
\eeq
for all  $x$ such that $|x-x_j|\leq\be c_n$. From this we obtain 
\begin{eqnarray}
\nn&&\sup_{x\in I_n}\left|\frac 1{nc_n}\sum_{i=1}^nK^{(\nu)}\left(\frac{x-X_{i-1}}{c_n}\right)X_i^k-E\left[\frac 1{nc_n}\sum_{i=1}^nK^{(\nu)}\left(\frac{x-X_{i-1}}{c_n}\right)X_i^k\right]\right|\\
\label{3.3-1}&\leq&\max_{1\leq j\leq M_n^*}\left|\frac 1{nc_n}\sum_{i=1}^nK^{(\nu)}\left(\frac{x_j-X_{i-1}}{c_n}\right)X_i^k-E\left[\frac 1{nc_n}\sum_{i=1}^nK^{(\nu)}\left(\frac{x_j-X_{i-1}}{c_n}\right)X_i^k\right]\right|\\
\label{3.3-2}&&{}+\max_{1\leq j\leq M_n^*} \be\left|\frac 1{nc_n}\sum_{i=1}^n \bK_I\left(\frac{x_j-X_{i-1}}{c_n}\right)\left|X_i^k\right|-E\left[\frac 1{nc_n}\sum_{i=1}^n \bK_I\left(\frac{x_j-X_{i-1}}{c_n}\right)\left|X_i^k\right|\right]\right|\qquad\quad\\
\nn&&{}+2\be\max_{1\leq j\leq M_n^*} E\left[\frac 1{nc_n}\sum_{i=1}^n \bK_I\left(\frac{x_j-X_{i-1}}{c_n}\right)\left|X_i^k\right|\right].
\end{eqnarray}
The last term on the right hand side can be bounded by
\beq
&&2\be\sup_{x\in I_n}E\left[\frac 1{nc_n}\sum_{i=1}^n \bK_I\left(\frac{x-X_{i-1}}{c_n}\right)\left|X_i^k\right|\right]\\
&=&2\be\sup_{x\in I_n}\frac 1{n}\sum_{i=1}^n \int \bK_I\left(u\right)E\left[\left|X_i^k\right|\bigg|X_{i-1}=x-uc_n\right]f_{X_{i-1}}(x-uc_n)\,du\\
&\leq&2\be\int  \bK_I\left(u\right) \,du\,\sup_{x\in I_n}\frac 1n\sum_{i=1}^nE\left[\left|X_i^k\right|\bigg|X_{i-1}=x\right]f_{X_{i-1}}(x)\;=\;O(\be)
\eeq
by a change of variable and by assumption (K) and (Z) or (Z'), respectively, and model assumption (AR).
In what follows we show that term (\ref{3.3-1}) is of order $O_P(\be)$; it can be shown analogously that (\ref{3.3-2}) is of the same order. Define (for $j$ fixed)
$$Y_{i}=K^{(\nu)}\left(\frac{x_j-X_{i-1}}{c_n}\right)X_i^k-E\left[K^{(\nu)}\left(\frac{x_j-X_{i-1}}{c_n}\right)X_i^k\right],$$
then the sequence $(Y_i)_i$ inherits the mixing conditions from $(X_i)_i$ due to 2.6.1 (ii) in Fan \& Yao (2005). Further the variables are centered and bounded by $2\bK n^{1/b}\log n$. We apply Liebscher's (1996) Theorem 2.1 to $\sum_{i=1}^n Y_i$ to obtain 
\begin{eqnarray}
\nn&&P\left(\max_{1\leq j\leq M_n^*}\left|\frac 1{nc_n}\sum_{i=1}^nK^{(\nu)}\left(\frac{x_j-X_{i-1}}{c_n}\right)X_i^k-E\left[\frac 1{nc_n}\sum_{i=1}^nK^{(\nu)}\left(\frac{x_j-X_{i-1}}{c_n}\right)X_i^k\right]\right|>\bM\be\right)\\
&\leq&M_n\left(4\exp\left(-\frac{nc_n\bM^2\be^2}{64\left(1+m_nc_n\right)A(m_n)+\frac {16}3\bM \bK m_nn^{\frac 1b}\be\log(n) }\right)+4\frac nm_n \alpha(m_n)\right) \label{liebscher}
\end{eqnarray}
for some $\bM$ independent of $j$ and 
for 
\begin{eqnarray}\label{m_n}
m_n&=&\left\{\begin{array}{ll} \lfloor n\be^2 (\log M_n)^{-1}(\log n)^{-1}\rfloor& \mbox{if } n^{\frac 1b-\frac 12}c_n^{-\frac 32}(\log n)^{\frac 52}=O(1)\\
\lfloor n^{1-\frac 1b}c_n\be (\log M_n)^{-1}(\log n)^{-2}\rfloor & \mbox{otherwise.}\end{array}
\right.
\end{eqnarray}
Further, 
\beq
A(m_n)&=&2(C+\be)\bK^2\sup_{x\in J_n}\max_{j^*+1\leq i\leq n-j^*}\sum_{j=i-j^*}^{i+j^*}\sqrt{E\left[X_j^{2k}\bigg|X_{j-1}=x\right]f_{X_{j-1}}(x)}\\
&&\qquad\times\sup_{x\in J_n}\frac 1{m_n}\max_{0\leq S\leq n-m_n}\sum_{i=S+1}^{S+m_n} \sqrt{E\left[X_i^{2k}\bigg|X_{i-1}=x\right]f_{X_{i-1}}(x)}\\
&&{}+\left(2(C+\be)\bK\right)^2\left(\left(\sup_{x\in J_n}\frac 1{m_n}\max_{0\leq S\leq n-m_n}\sum_{i=S+1}^{S+m_n} E\left[|X_i|^k\bigg|X_{i-1}=x\right]f_{X_{i-1}}(x)\right)^2\right.\\
&&\qquad +\left.\sup_{x,x'\in J_n}\!\frac 1{m_n^2}\max_{0\leq S\leq n-m_n}\!\!\!\sum_{\substack{i,j=S+1\\ |i-j|>j^*}}^{S+m_n}\!\!\!\! E\!\left[|X_i|^k|X_j|^k\bigg|X_{i-1}=x,X_{j-1}=x'\right]f_{X_{i-1},X_{j-1}}(x,x')\right),
\eeq
with $j^*$ as in assumption (Z) resp.\ (Z'). In order to obtain \ref{liebscher} from Liebscher's Theorem one has to show that 
\beq
\max_{0\leq T\leq n-1}E\left[\left(\sum_{i=T+1}^{\min(T+m_n,n)} Y_i\right)^2\right]
&\leq&\left(m_nc_n+m_n^2c_n^2\right)A(m_n).
\eeq
This can be done by some tedious calculations, which are omitted for the sake of brevity. 

By assumption (Z) resp.\ (Z') we have $A(m_n)=O(1)$. To see this consider for example for $k=1$ the term 
\beq
&&\sup_{x\in J_n}\frac 1{m_n}\max_{0\leq S\leq n-m_n}\sum_{i=S+1}^{S+m_n} E\left[|X_i|\bigg|X_{i-1}=x\right]f_{X_{i-1}}(x)\\
&=&\sup_{x\in J_n}\frac 1{m_n}\max_{0\leq S\leq n-m_n}\sum_{i=S+1}^{S+m_n} E\left[|m(x)+\sigma(x)\e_i|\right]f_{X_{i-1}}(x)\\
&\leq&\sup_{x\in J_n}\left(\left(|m(x)|+2|\sigma(x)|\right)\ \frac 1{m_n}\max_{0\leq S\leq n-m_n}\sum_{i=S+1}^{S+m_n}f_{X_{i-1}}(x)\right)\;=\;O(1)
\eeq
(note that $m_n^{-1}=o(1)$), and
\beq
&&\sup_{x\in J_n}\frac 1{m_n}\max_{0\leq S\leq n-m_n}\sum_{i=S+1}^{S+m_n} \sqrt{E\left[X_i^2\bigg|X_{i-1}=x\right]f_{X_{i-1}}(x)}\\
&\leq&\sup_{x\in J_n}\frac 1{m_n}\max_{0\leq S\leq n-m_n}\left(m_n
+\sum_{\substack{i=S+1\\ E\left[X_i^2\bigg|X_{i-1}=x\right]f_{X_{i-1}}(x)> 1}}^{S+m_n} E\left[X_i^2\bigg|X_{i-1}=x\right]f_{X_{i-1}}(x)\right)\\
&\leq& 1+\sup_{x\in J_n}\frac 1{m_n}\max_{0\leq S\leq n-m_n}\sum_{i=S+1}^{S+m_n} E\left[X_i^2\bigg|X_{i-1}=x\right]f_{X_{i-1}}(x)\\
&\leq&1+\sup_{x\in J_n}\left(\left( |m(x)|+|\sigma(x)|\right)^2\ \frac 1{m_n}\max_{0\leq S\leq n-m_n}\sum_{i=S+1}^{S+m_n} f_{X_{i-1}}(x)\right)\;=\;O(1).
\eeq
The other terms in the definition of $A(m_n)$ are treated similarly. 

Inserting the definitions of $M_n$, $\be$, $m_n$ and $A(m_n)=O(1)$ one obtains with a simply calculation that (\ref{liebscher}) is of order $o(1)$ by the assumptions on the bandwidth $c_n$ and the mixing coefficient. This concludes the proof. 
\hfill $\Box$

\medskip

\noindent {\bf Proof of Lemma \ref{lemma3.4}.} 
We only present the proofs for the assertions on $\hat m$, those on $\hat\sigma$ follow by similar arguments using 
(\ref{sig-dar}).

Let $\hat{g}_k(x)=\frac 1{nc_n}\sum_{i=1}^nK\left(\frac{x-X_{i-1}}{c_n}\right)X_i^k$ for $k\in\{0,1,2\}$. Then from Lemma \ref{lemma3.3} it directly follows that
\begin{eqnarray}\label{3.4-1}
\sup_{x\in I_n}\left|\frac{\partial^{\nu}}{\partial x^{\nu}}\biggl(\hat{g}_k(x)-E\left[\hat{g}_k(x)\right]\biggr)\right| = O_P\left(\frac{\be}{c_n^\nu}\right)=O_P\left(c_n^{-\frac 12-\nu}n^{-\frac 12}(\log n)^{\frac 12}\right).
\end{eqnarray}
Let further $g_{k,i}(x)=E\left[X_i^k|X_{i-1}=x\right]f_{X_{i-1}}(x)$ for $i=1,\ldots,n$. Then 
\beq
&&\sup_{x\in I_n}\left|\frac{\partial^{\nu}}{\partial x^{\nu}}\left(E\left[\hat{g}_k(x)\right]-\frac 1n\sum_{i=1}^ng_{k,i}(x)\right)\right|\\
&=&\sup_{x\in I_n}\left|\frac 1{c_n^{\nu}}\frac 1{nc_n}\sum_{i=1}^n\int K^{(\nu)}\left(\frac{x-y}{c_n}\right)E\left[X_i^k\bigg|X_{i-1}=y\right]f_{X_{i-1}}(y)dy-\frac 1n\sum_{i=1}^ng^{(\nu)}_{k,i}(x)\right|\\
&=& \sup_{x\in I_n}\left|\frac 1{c_n^{\nu}}\frac 1{n}\sum_{i=1}^n\int K^{(\nu)}\left(u\right)g_{k,i}(x-uc_n)du-\frac 1n\sum_{i=1}^ng^{(\nu)}_{k,i}(x)\right|.
\eeq
Noting that by assumption (K),
\[\int  K^{(\nu)}\left(u\right)u^{\mu}du=0\text{ for }\  \mu\leq\nu+1,\  \mu\neq\nu,\quad 
\int  K^{(\nu)}\left(u\right)u^{\nu}du=(-1)^{\nu}\nu!\]
for $\nu\in\{0,1,2\}$,
from a Taylor expansion it further follows that
\begin{eqnarray}
\nn&&\sup_{x\in I_n}\left|\frac{\partial^{\nu}}{\partial x^{\nu}}\left(E\left[\hat{g}_k(x)\right]-\frac 1n\sum_{i=1}^ng_{k,i}(x)\right)\right|\\
&\leq&\frac{c_n^2}{(\nu+2)!}\sup_{x\in J_n}\left| \frac 1n\sum_{i=1}^ng^{(\nu+2)}_{k,i}(x)\right|\left|\int  K^{(\nu)}\left(u\right)u^{\nu+2}du\right| \;=\; O(c_n^2 q_n^k), \label{3.4-2}
\end{eqnarray}
where the last equality follows from assumption (X) resp.\ (X') and (M) (note that $g_{0,i}=f_{X_{i-1}}$, $g_{1,i}=mf_{X_{i-1}}$, $g_{2,i}=(m^2+\sigma^2)f_{X_{i-1}}$).
Note further that 
\begin{eqnarray}
\nn&&\inf_{x\in I_n}\left|\frac{\hat{g}_0(x)}{\frac 1n\sum_{i=1}^ng_{0,i}(x)}\right|
\;=\;\inf_{x\in I_n}\left|1+\left(\frac{\hat{g}_0(x)-\frac 1n\sum_{i=1}^ng_{0,i}(x)}{\frac 1n\sum_{i=1}^ng_{0,i}(x)}\right)\right|\\
&\geq&1-\frac{\sup_{x\in I_n}\left|\hat{g}_0(x)-\frac 1n\sum_{i=1}^ng_{0,i}(x)\right|}{\inf_{x\in I_n}\left|\frac 1n\sum_{i=1}^nf_{X_{i-1}}(x)\right|}\;=\;
1+O_P\left(c_n^{-\frac 12}n^{-\frac 12}(\log n)^{\frac 12}q_n^f+c_n^2q_n^f\right) \quad\label{3.4-3}
\end{eqnarray}
by (\ref{3.4-1}), (\ref{3.4-2}) and assumption (X) resp.\ (X').

Now for $\hat m=\hat g_1/\hat g_0$ from (\ref{nw-m}) we obtain
\beq
&&\sup_{x\in I_n}|\hat{m}(x)-m(x)|\\
&=&\sup_{x\in I_n}\left| \frac{\hat{g}_1(x)-\frac 1n\sum_{i=1}^ng_{1,i}(x)+m(x)\left(\frac 1n\sum_{i=1}^ng_{0,i}(x)-\hat{g}_0(x)\right)}{\hat{g}_0(x)}\right|\\
&=&\sup_{x\in I_n}\left| \frac{\frac1{\frac 1n\sum_{i=1}^nf_{X_{i-1}}(x)}\left(\hat{g}_1(x)-\frac 1n\sum_{i=1}^ng_{1,i}(x)+m(x)\left(\frac 1n\sum_{i=1}^ng_{0,i}(x)-\hat{g}_0(x)\right)\right)}{\frac{\hat{g}_0(x)}{\frac 1n\sum_{i=1}^ng_{0,i}(x)}}\right|\\
&\leq&\frac{\frac 1 {\inf_{x\in I_n}\frac 1n\sum_{i=1}^nf_{X_{i-1}}(x)}}{\inf_{x\in I_n}\left|\frac{\hat{g}_0(x)}{\frac 1n\sum_{i=1}^ng_{0,i}(x)}\right|}\left(\sup_{x\in I_n}\!\left|\hat{g}_1(x)\!-\!\frac 1n\sum_{i=1}^ng_{1,i}(x)\right|+\sup_{x\in I_n}\!|m(x)|\!\sup_{x\in I_n}\!\left|\frac 1n\sum_{i=1}^ng_{0,i}(x)\!-\!\hat{g}_0(x)\right|\right)\\
&=&O_P\left(c_n^{-\frac 12}n^{-\frac 12}(\log n)^{\frac 12}q_nq_n^f+c_n^2q_nq_n^f\right)
\eeq
by (\ref{3.4-1}), (\ref{3.4-2}), (\ref{3.4-3}) and assumption (X) resp.\ (X'). The first assertion in (i) now directly follows from 
$(\inf_{x\in I_n}|\sigma(x)|)^{-1}=O\left(q_n^{\sigma}\right)$ (assumption (M)).

Differentiating it is easy to see that from (\ref{3.4-1}) and (\ref{3.4-2}) it follows that
\begin{eqnarray*}
&&\sup_{x\in I_n}\left|\frac{\partial^{\nu}}{\partial x^{\nu}}\left(\frac{\hat{m}(x)-m(x)}{\sigma(x)}\right)\right|\\
&=&O\left((q_n^{\sigma})^{\nu+1}\right)\sum_{j=0}^{\nu}O\left((q_n^f)^{j+1}\right)\sum_{l=0}^1 O\left(q_n^{1-l}\right)\cdot O\left(\sup_{x\in I_n}\left|\frac{\partial^{j}}{\partial x^{j}}\left(\hat{g}_l(x)-\frac 1n\sum_{i=1}^ng_{l,i}(x)\right)\right|\right)\\
&=&O_P\left(\left(c_n^{-\frac 12-\nu}n^{-\frac 12}(\log n)^{\frac 12}+c_n^2\right)q_n(q_n^fq_n^{\sigma})^{\nu+1}\right)
\;=\; o_P(1),
\end{eqnarray*}
where the last equality only holds for $\nu=0,1$
by our bandwidth conditions. The first assertion of (ii) follows. 

Finally, note that by considering the cases $|x-y|\leq c_n$ and $|x-y|> c_n$ we have
\beq
&&\sup_{x,y\in I_n,x\neq y}\frac{\left|\frac{\partial}{\partial x}\left(\frac{\hat{m}(x)-m(x)}{\sigma(x)}\right)-\frac{\partial}{\partial y}\left(\frac{\hat{m}(y)-m(y)}{\sigma(y)}\right)\right|}{|y-x|^{\delta}}\\
&\leq&2\cdot \sup_{x\in I_n}\left|\frac{\partial}{\partial x}\left(\frac{\hat{m}(x)-m(x)}{\sigma(x)}\right)\right|c_n^{-\delta}+\sup_{x\in I_n}\left|\frac{\partial^2}{\partial x^2}\left(\frac{\hat{m}(x)-m(x)}{\sigma(x)}\right)\right|\sup_{x,y\in I_n,0<|x-y|\leq c_n}|x-y|^{1-\delta}\\
&=&O_P\left(\left(c_n^{-\frac 32}n^{-\frac 12}(\log n)^{\frac 12}+c_n^2\right)q_n(q_n^fq_n^{\sigma})^2\right)c_n^{-\delta}+O_P\left(\left(c_n^{-\frac 52}n^{-\frac 12}(\log n)^{\frac 12}+c_n^2\right)q_n(q_n^fq_n^{\sigma})^3\right)c_n^{(1-\delta)}\\
&=&o_P(1)
\eeq
and the first assertion of (iii) follows. 
\hfill $\Box$

\medskip

\noindent {\bf Proof of Lemma \ref{lemma-10}.} 
{\bf (i).} For 
\[\hat{d}_n(x)= w_n(x)\frac 1n\sum_{i=1}^n\e_i\left(\frac{\frac 1{c_n}K\left(\frac{x-X_{i-1}}{c_n}\right)-\frac 1{nc_n}\sum_{k=1}^nK\left(\frac{x-X_{k-1}}{c_n}\right)}{\frac 1{nc_n}\sum_{k=1}^nK\left(\frac{x-X_{k-1}}{c_n}\right)}\right)\]
we have
\begin{eqnarray*}
 \int \hat{d}_n(x)f_{X_{0}}(x)\,dx&=& \frac 1n\sum_{i=1}^n Z_i \int w_n(x)\left(\frac 1{c_n}K\left(\frac{x-X_{i-1}}{c_n}\right)-f_{X_0}(x)\right)\,dx+\op\\
&=&\op,
\end{eqnarray*}
where last equality follows by a calculation of the variance and Chebyshev's inequality. The first equality can be derived by using
\begin{eqnarray}\label{dichterate}
\sup_{x\in I_n}\left|\frac 1{nc_n}\sum_{k=1}^nK\left(\frac{x-X_{k-1}}{c_n}\right)-f_{X_{0}}(x)\right|&=&O_P\left(c_n^{-\frac 12}n^{-\frac 12}(\log n)^{\frac 12}+c_n^2\right),
\end{eqnarray}
which follows from the proof of Lemma \ref{lemma3.4} (note that with the notations used there, $(nc_n)^{-1}\sum_{k=1}^nK((x-X_{k-1})/c_n)=\hat g_0(x)$, $f_{X_0}(x)=n^{-1}\sum_{i=1}^n g_{0,i}(x)$) and results from Lemma \ref{lemma3.4} (i) for the AR-model with $m\equiv 0$, $\sigma\equiv 1$.

Now assertion (i) is equivalent to 
\begin{equation*}
\sup_{s\in[0,1]}\left|\frac 1n\sjns \tilde{d}_n(X_{j-1})\right|=\op.
\end{equation*}
with centered functions $\tilde d_n=\hat d_n-\int \hat d_n(x) f_{X_0}(x)\,dx$. By arguments similar to those in the proof of Lemma \ref{lemma3.4} and by results from Lemma \ref{lemma3.4} for $m\equiv 0$, $\sigma\equiv 1$ we have $P(\tilde d_n\in\mathcal{D}_n)\to 1$ for $n\to\infty$ for the function classes
\begin{eqnarray*}
\mathcal{D}_n&=&\Big\{d:I_n\to \er\Big|\max\left(\sup_{x\in I_n}|d(x)|,\sup_{x\in I_n}|d'(x)|\right)+\sup_{x,y\in I_n, x\neq y}\farc{|d'(x)-d'(y)|}{|x-y|^{\delta}}\leq 1,\\ 
&&\quad\sup_{x\in I_n}|d(x)|\leq z_n,
 \int d(y)f_{X_0}(y)dy=0\Big\}
\end{eqnarray*}
with $z_n=c_n^{-\frac 12}n^{-\frac 12}(\log n)^{\frac 12}q_n^f(\log n)^{\frac 12}$. Thus it remains to show
that
\begin{equation}\label{cpbewentwnwlemmaK21}
\sup_{s\in[0,1]}\sup_{d\in \mathcal{D}_n}\left|\frac 1n\sjns d(X_{j-1})\right|=\op.
\end{equation}
To this end let $\be=n^{-\frac 12}(\log n)^{-1}$. It follows from Theorem 2.7.1 in van der Vaart \& Wellner (1996) that 
$M_{n2}\leq \exp(2^{\frac1{1+\delta}}\bK(2+b_n-a_n)\bar{\e}_n^{-\frac 1{1+\delta}})$ balls of radius $\be$ with respect to the supremum norm $||\cdot||_{I_n}$ on the interval $I_n$ are needed to cover $\mathcal{D}_n$. Here the constant $\bK$ only depends on $\delta$.
Let $d_{1},\ldots,d_{M_{n2}}$ denote centers of those balls. We may assume that those functions are elements of $\mathcal{D}_n$, too (see Pollard (1990), p.\ 10). 
Further let $0=s_1<\ldots<s_{M_{n1}}=1$ segment $[0,1]$ in intervals of length $\leq\be/z_n$ such that $M_{n1}\leq \frac{z_n}{\be}$. Then it can be shown that
\beq
&&\sup_{s\in[0,1]}\ \sup_{d\in \mathcal{D}_n} \left|\frac 1n\sjns d(X_{j-1})\right|\\
&\leq&\max_{1\leq h\leq M_{n1}}\ \max_{1\leq k\leq M_{n2}} \left|\frac 1n\sjnsh d_{k}(X_{j-1})\right|\\
&&{}+\max_{1\leq h\leq M_{n1}}\sup_{s\in[0,1]\ \text{mit}\ |s-s_h|\leq \be/z_n}\ \sup_{d\in \mathcal{D}_n} \left|\frac 1n\sjn d(X_{j-1})\left(I\left\{\frac jn\leq s\right\}-I\left\{\frac jn\leq s_h\right\}\right)\right|\\
&&{}+\max_{h}\ \max_{1\leq k\leq M_{n2}}\sup_{d\in \mathcal{D}_n\ \text{mit}\ \|d-d_{k}\|_{I_n}\leq\be} \left|\frac 1n\sjnsh \left(d(X_{j-1})-d_{k}(X_{j-1})\right)\right|\\
&\leq&\max_{1\leq h\leq M_{n1}}\ \max_{1\leq k\leq M_{n2}}\ \left|\frac 1n\sjnsh d_{k}(X_{j-1})\right|+o\left(\frac 1{\sqrt n}\right).
\eeq
By an application of Liebscher's (1996) Theorem 2.1 to random variables $Y_i=d_k(X_{i-1})I\left\{\frac in\leq s_h\right\}$ (for $k,h$ fixed) one can show the existence of some constant $\bM$ such that
\begin{eqnarray*}
&&P\left(\max_{1\leq h\leq M_{n1}}\ \max_{1\leq k\leq M_{n2}}\ \left|\frac 1n\sjnsh d_{k}(X_{j-1})\right|>\bM\be\right)\\
&\leq&\sum_{h=1}^{M_{n1}}\sum_{k=1}^{M_{n2}}P\left(\left|\frac 1n\sjnsh d_{k}(X_{j-1})\right|>\bM\be\right)\\
&\leq&M_{n1}M_{n2}\left(4\exp\left(-\frac{n^2\bM^2\be^2}{64n\lfloor n\be c_n^{\frac 12}\rfloor z_n^2+\frac 83n\bM\be \lfloor n\be c_n^{\frac 12}\rfloor z_n}\right)+4\frac n{\lfloor n\be c_n^{\frac 12}\rfloor}\alpha(\lfloor n\be c_n^{\frac 12}\rfloor)\right)\\
&=& o(1).
\end{eqnarray*}
Details are omitted for the sake of brevity. From this the rate $O_P(\be)=o_P(n^{-1/2})$ follows for (\ref{cpbewentwnwlemmaK21}).

\noindent{\bf (ii).} We only describe the main steps of this proof. The random denominator can be replaced by the true density $f_{X_0}$ due to (\ref{dichterate}). Now define
$$\hat d_n(x)=w_n(x)\frac{1}{nc_n}\sum_{i=1}^n K\left(\frac{x-X_{i-1}}{c_n}\right)\e_i\frac{\sigma(X_{i-1})-\sigma(x)}{\sigma(x)f_{X_0}(x)}.$$
Then, $\int \hat d_n(x)f_{X_0}(x)\,dx=o_P(n^{-1/2})$ can be shown by Chebyshev's inequality and for $\tilde d_n=\hat d_n-\int \hat d_n(x)f_{X_0}(x)\,dx$ the assertion 
\begin{equation*}
\sup_{s\in[0,1]}\left|\frac 1n\sjns \tilde{d}_n(X_{j-1})\right|=\op.
\end{equation*}
is shown analogously to the proof of (i). 

\noindent{\bf (iii).} The assertion can be proved by the same methods. 
\hfill $\Box$

\medskip

\noindent {\bf Proof of Lemma \ref{lemma9.1}.} 
Let $\hat{d}_{n1}=(\hat m-m)/\sigma$ and $\hat{d}_{n2}=\hat{\sigma}/\sigma$.
Now the assumption of the lemma is equivalent to 
\[\sup_{s\in [0,1]}\sup_{t\in\er}|H_n(s,t,\hat{d}_{n1},\hat{d}_{n2})-H_n(s,t,0,1)|=o_P(\frac{1}{\sqrt{n}})\]
with
\[H_n(s,t,d_1,d_2)=\frac 1n\sum_{j=1}^{\ns}w_{nj}\left(I\{\epsilon_j\leq t\cdot d_{2}(X_{j-1})+d_{1}(X_{j-1})\}-F_{\e_j}(td_{2}(X_{j-1})+d_{1}(X_{j-1}))\right).\]
For the proof we may assume that 
$\sup_{x\in I_n}|\hat{d}_{n1}(x)|<1$, $\inf_{x\in I_n}\hat{d}_{n2}(x)>\frac 12$ and
$|\e_j|\leq\sqrt n\log n$ for all $j=1,\ldots,n$  because
$\sup_{x\in I_n}|\hat{d}_{n1}(x)|=o_P(1)$, $\sup_{x\in I_n}|\hat{d}_{n2}(x)-1|=o_P(1)$ by Lemma \ref{lemma3.4}, and further\begin{eqnarray*}
P\Big(\max_{1\leq j\leq n}|\e_j|>\sqrt n\log n\Big)
&\leq& \frac 1{n\log n}\sjn E\Big[\e_j^2I\Big\{\e_j^2>n(\log n)^2\Big\}\Big]
\;\leq\;\frac 1{\log n }\;=\;o(1)
\end{eqnarray*}
by the model assumption  $E[\e_j^2]=1$ for all $j\in\mathbb{Z}$.

Note that $I\{\varepsilon_j\leq t\cdot \hat{d}_{n2}(X_{j-1})+\hat{d}_{n1}(X_{j-1})\}-I\{\varepsilon_j\leq t\}=0$
for all $t$ such that $|t|>\sqrt n\log n$, for all $j=1,\ldots,n$. Thus, by some simple estimations, 
\begin{eqnarray*}
&&\sup_{s\in [0,1]}\sup_{ |t|>\sqrt n\log n}|H_n(s,t,\hat{d}_{n1},\hat{d}_{n2})-H_n(s,t,0,1)|\\
&\leq&\sup_{s\in [0,1]}\sup_{| t|>\sqrt n\log n}\frac 1n\sum_{j=1}^{\ns}w_{nj}\left|-F_{\e_j}(t\cdot \hat{d}_{n2}(X_{j-1})+\hat{d}_{n1}(X_{j-1}))+F_{\e_j}(t)\right|\\
&\leq& \frac 1n\sum_{j=1}^n\left(1-F_{\e_j}\left(\frac{\sqrt n\log n}2-1\right)\right)\ +\ \frac 1n\sum_{j=1}^nF_{\e_j}\left(-\frac{\sqrt n\log n}2+1\right)\\
&=& O\left(\frac 1{n(\log n)^2}\right)\;=\;o\left(\frac 1{\sqrt n}\right),
\end{eqnarray*}
where in the last line we have applied that for $t>1$
\begin{eqnarray}\label{9.1t^2}
\frac 1n\sjn (1-F_{\e_j}(t))&=&\frac 1n\sjn P(\e_j>t)\;\leq\; \frac 1{t^2}\frac 1n\sjn E[\e_j^2I\{\e_j^2>t^2\}]\;=\;O\left(\frac 1{t^2}\right)
\end{eqnarray}
by the model assumption $E[\e_j^2]=1$ and analogously for $t<-1$, $\frac 1n\sjn F_{\e_j}(t)=O(1/t^2)$.

For the remainder of the proof we therefore only need to consider $|t|\leq \sqrt{n}\log n$. Define sequences of function classes by
\begin{eqnarray*}
\mathcal{D}_{1,n}&=&\Bigg\{d:I_n\to\er\Bigg|\max\{\sup_{x\in I_n}|d(x)|,\sup_{x\in I_n}|d'(x)|\}+\sup_{x,y\in I_n,x\neq y}\frac{|d'(x)-d'(y)|}{|y-x|^{\delta}}\leq 1,\\
&&\quad\sup_{x\in I_n}|d(x)|\leq z_{n1}\log n\Bigg\} \mbox{ for } z_{n1}=\left(c_n^{-\frac 12}n^{-\frac 12}(\log n)^{\frac 12}+c_n^2\right)q_nq_n^fq_n^{\sigma}\\
\mathcal{D}_{2,n}&=&\Bigg\{d:I_n\to\er\Bigg|\max\{\sup_{x\in I_n}|d(x)|,\sup_{x\in I_n}|d'(x)|\}+\sup_{x,y\in I_n,x\neq y}\frac{|d'(x)-d'(y)|}{|y-x|^{\delta}}\leq 2,\\
&&\quad\inf_{x\in I_n}d(x)\geq\frac 12, \sup_{x\in I_n}|d(x)-1|\leq z_{n2}\log n\Bigg\} \mbox{ for } z_{n2}=\left(c_n^{-\frac 12}n^{-\frac 12}(\log n)^{\frac 12}+c_n^2\right)(q_nq_n^fq_n^{\sigma})^2.
\end{eqnarray*}
Then by Lemma \ref{lemma3.4}, $P(\hat d_{n1}\in \mathcal{D}_{1,n})\to 1$, $P(\hat d_{n2}\in \mathcal{D}_{2,n})\to 1$ and it remains to show that
\[\sup_{s\in[0,1],|t|\leq\sqrt n\log n,\atop d_1\in \mathcal{D}_{1,n},d_2\in\mathcal{D}_{2,n}}\left|H_n(s,t,d_1,d_2)-H_n(s,t,0,1)\right|=o_p\left(\frac{1}{\sqrt n}\right).\]
To this end we apply covering arguments. Let $\be=\min\{\frac 18,n^{-\frac 12}(\log n)^{-1}\}$. The $\be$-covering numbers  of both function classes with respect to the supremum norm on $I_n$ can be bounded by $M_n\leq\exp(c(2+b_n-a_n)\be^{-1/(1+\delta)})$, see Theorem 2.7.1 by van der Vaart \& Wellner (1996). 
Let $d_{11},\ldots,d_{1M_{n}}$ and $d_{21},\ldots,d_{2M_{n}}$, respectively, denote the corresponding centers of covering balls. Note that then $\sup_{x\in I_n}|d_{1k}(x)|\leq 1+\be$ and $d_{2l}(x)\in\left[\frac 12-\be,2+\be\right]$ for all $x\in I_n$. Let further the intervals $[0,1]$ and $[-\sqrt n\log n,\sqrt n\log n]$ be segmented by points $0= s_1<\ldots <s_{M_{ns}}=1$ and $-\sqrt n\log n= t_1<\ldots <t_{M_{nt}}=\sqrt n\log n$, respectively, in segments of length $\leq \bar{\e}_n$ such that the number of points are bounded by $M_{ns}\leq 1/ \bar{\e}_n$ and $M_{nt}\leq 2\sqrt n\log n /\bar{\e}_n$. Let $||\cdot||_{I_n}$ denote the supremum norm on $I_n$. Then 
\begin{eqnarray}
\nn&&\sup_{s\in[0,1],|t|\leq\sqrt n\log n,\atop d_1\in \mathcal{D}_{1,n},d_2\in\mathcal{D}_{2,n}}\left|H_n(s,t,d_1,d_2)-H_n(s,t,0,1)\right|\\
\nn&&\\
\nn&\leq& \max_{h,i,k,l}\left|H_n(s_h,t_i,d_{1k},d_{2l})-H_n(s_h,t_i,0,1)\right|\\
\nn&&\\
\label{cpbewlemma6}&\ &+\max_{h}\sup_{|s-s_h|\leq\be,|t|\leq \sqrt n\log n,\atop d_1\in \mathcal{D}_{1,n},d_2\in\mathcal{D}_{2,n}}|H_n(s,t,d_1,d_2)-H_n(s_h,t,d_{1}, d_{2})|\\
\label{cpbewlemma3}&\ &+\max_{h,i,k,l}\sup_{|t-t_i|\leq\be,\atop \|d_1-d_{1k}\|_{I_n}\leq\be,\|d_2-d_{2l}\|_{I_n}\leq\be}|H_n(s_h,t,d_1,d_2)-H_n(s_h,t_i,d_{1k}, d_{2l})|\\
\nn&&\\
\label{cpbewlemma4}&\ &+\max_{h}\sup_{|s-s_h|\leq\be,|t|\leq \sqrt n\log n}|H_n(s_h,t,0,1)-H_n(s,t,0,1)|\\
\label{cpbewlemma5}&\ &+\max_{h,i}\sup_{|t-t_i|\leq\be}|H_n(s_h,t_i,0,1)-H_n(s_h,t,0,1)|,
\end{eqnarray}
where the maximum is always with respect to $h\in\{1,\ldots,M_{ns}\}$, $i\in\{1,\ldots,M_{nt}\}$, $k,l\in\{1,\ldots,M_{n}\}$.

To further bound the term (\ref{cpbewlemma3}) first consider fixed $h\in\{1,\ldots,M_{ns}\}$, $i\in\{1,\ldots,M_{nt}\}$, $k,l\in\{1,\ldots,M_{n}\}$ such that $t_i\geq\be$ (the other case is treated analogously). Then
\begin{eqnarray*}
&&\sup_{|t-t_i|\leq\be,\atop \|d_1-d_{1k}\|_{I_n}\leq\be,\|d_2-d_{2l}\|_{I_n}\leq\be}|H_n(s_h,t,d_1,d_2)-H_n(s_h,t_i,d_{1k}, d_{2l})|\\
&\leq &\!\!\!\sup_{|t-t_i|\leq\be,\atop \|d_1-d_{1k}\|_{I_n}\leq\be,\|d_2-d_{2l}\|_{I_n}\leq\be}\!\!\!  \frac 1n\sjn w_{nj}|I\{\varepsilon_j\leq td_{2}(X_{j-1})+d_{1}(X_{j-1})\}-I\{\varepsilon_j\leq td_{2l}(X_{j-1})+d_{1k}(X_{j-1})\}|\\
&&{}+\!\!\!\sup_{|t-t_i|\leq\be,\atop \|d_1-d_{1k}\|_{I_n}\leq\be,\|d_2-d_{2l}\|_{I_n}\leq\be}\!\!\!  \frac 1n\sjn w_{nj}|F_{\e_j}(td_{2l}(X_{j-1})+d_{1k}(X_{j-1}))-F_{\e_j}(td_{2}(X_{j-1})+d_{1}(X_{j-1}))|\\
&\leq & H_n(1,t_i+\be,d_{1k}+\be, d_{2l}+\be)-H_n(1,t_i-\be,d_{1k}-\be, d_{2l} -\be)\\
&& {}+ \frac 2n\sjn w_{nj}\Bigg(F_{\e_j}((t_i+\be)(d_{2l}(X_{j-1})+\be)+d_{1k}(X_{j-1})+\be)\\
&&\qquad{}-F_{\e_j}((t_i-\be)(d_{2l}(X_{j-1})-\be)+d_{1k}(X_{j-1})-\be)\Bigg)\\
&=&H_n(1,t_i+\be,d_{1k}+\be, d_{2l}+\be)-H_n(1,t_i-\be,d_{1k}-\be, d_{2l} -\be) +o\left(\frac 1{\sqrt n}\right),
\end{eqnarray*}
where the last step follows from the mean value theorem, assumption (F) resp.\ (F') and $\be=o(1/\sqrt n)$. 
Similarly for (\ref{cpbewlemma5}) we obtain
\begin{eqnarray}
\nn&&\max_{h,i}\sup_{|t-t_i|\leq\be}|H_n(s_h,t_i,0,1)-H_n(s_h,t,0,1)|\\
\label{cpbewlemma9}&\leq&\max_i |H_n(1,t_i+\be,0,1)-H_n(1,t_i-\be,0,1)|+o\left(\frac 1{\sqrt n}\right).
\end{eqnarray}
For (\ref{cpbewlemma6}) we have
\begin{eqnarray*}
\nn&&\max_{h}\sup_{|s-s_h|\leq\be,|t|\leq \sqrt n\log n,\atop d_1\in \mathcal{D}_{1,n},d_2\in\mathcal{D}_{2,n}}|H_n(s,t,d_1,d_2)-H_n(s_h,t,d_{1}, d_{2})|\\
\nn&\leq&\max_{h}\sup_{|s-s_h|\leq\be}\frac 1n\sjn\left|I\left\{\frac jn\leq s\right\}-I\left\{\frac jn\leq s_h\right\}\right|\\
\nn&&\\
\nn&\leq&\max_{h}\sup_{|s-s_h|\leq\be}\left(|s-s_h|+\frac 1n\right)
\;\leq\;\be+\frac 1n\;=\;o\left(\frac 1{\sqrt n}\right)
\end{eqnarray*}
and analogously for (\ref{cpbewlemma4}) the same rate $o(1/\sqrt{n})$. 
Altogether we have shown that
\begin{eqnarray}\nn
&&\sup_{s\in[0,1],|t|\leq\sqrt n\log n,\atop d_1\in \mathcal{D}_{1,n},d_2\in\mathcal{D}_{2,n}}
\left|H_n(s,t,d_1,d_2)-H_n(s,t,0,1)\right|\\
&\leq& \max_{h,i,k,l}\left|H_n(s_h,t_i,d_{1k},d_{2l})-H_n(s_h,t_i,0,1)\right|\label{freed}\\
&&{}+\max_{i,k,l}|H_n(1,t_i+\be,d_{1k}+\be,d_{2l}+\be)-H_n(1,t_i-\be,d_{1k}-\be, d_{2l}-\be)|\nn\\
&&{}+\max_i |H_n(1,t_i+\be,0,1)-H_n(1,t_i-\be,0,1)|+o(\frac 1{\sqrt n}).\nn
\end{eqnarray}
To conclude the proof we exemplarily consider term (\ref{freed}); the other terms are treated analogously. For all $\eta>0$ we have
 \begin{eqnarray}
\nn&& P\left(\sqrt n\max_{h,i,k,l}\left|H_n(s_h,t_i,d_{1k},d_{2l})-H_n(s_h,t_i,0,1)\right|>\eta\right)\\
\nn&\leq&\sum_{h=1}^{M_{ns}}\sum_{i=1}^{M_{nt}}\sum_{k=1}^{M_{n}}\sum_{l=1}^{M_{n}}P\left(\sqrt n\left|H_n(s_h,t_i,d_{1k},d_{2l})-H_n(s_h,t_i,0,1)\right|>\eta\right)\\
\label{cpbewlemma15}&\leq& M_{ns}\sum_{i=1}^{M_{nt}}\sum_{k=1}^{M_{n}}\sum_{l=1}^{M_{n}}\left(2\exp\left(-\frac{\eta^2n}{4\eta\sqrt n+2nA_{i,k,l}(n)}\right)\right)
\end{eqnarray}
by an application of Theorem 1.6 by Freedman (1975). To see this define (for $h,i,k,l$ fixed)
\begin{eqnarray*}
Y_j&=&w_{nj}I\left\{\frac jn\leq s_h\right\}\left(I\{\e_j\leq t_id_{2l}(X_{j-1})+d_{1k}(X_{j-1})\}-I\{\e_j\leq t_i\}\right.\\
&&\left.\hspace{3,5cm}-F_{\e_j}(t_id_{2l}(X_{j-1})+d_{1k}(X_{j-1}))+F_{\e_j}(t_i)\right),
\end{eqnarray*}
and note that  $|Y_j|\leq 2$, $E[Y_j|X_0,\ldots,X_{j-1}]=0=E[-Y_j|X_0,\ldots,X_{j-1}]$ as well as
\begin{eqnarray*}
\sum_{j=1}^nE\left[Y_j^2|X_0,\ldots,X_{j-1}\right] &\leq& nA_{i,k,l}(n)\;=\; \sum_{j=1}^n\sup_{x\in I_n}|F_{\e_j}(t_id_{2l}(x)+d_{1k}(x))-F_{\e_j}(t_i)|\\
&\leq & C n( z_{n1}\log n +z_{n2}\log n +\be )
\end{eqnarray*}
by an application of the mean value theorem, the definition of $\mathcal{D}_{1,n}$, $\mathcal{D}_{2,n}$ and assumption (F) resp.\ (F'),
for some constant $C$ independent of $i,k,l$. Hence, inserting the bounds on $M_n$, $M_{ns}$ and $M_{nt}$, $(\ref{freed})$ can be bounded by 
\beq
2\exp\Bigg(\log\left(\frac {\sqrt n\log n}{\be^2}\right)+2c(2+b_n-a_n)\bar{\e}_n^{-\frac 1{1+\delta}}\quad&&\\
{}-\frac{n\eta^2}{4\eta\sqrt n+2Cn(z_{n1}\log n+z_{n2}\log n+\be)}\Bigg)&=& o(1),
\eeq
which follows from bandwidth condition (\ref{delta}). 
\hfill $\Box$

\medskip

\noindent {\bf Proof of Lemma \ref{lemma9.3}.} 
Analogous to the proof of Lemma \ref{lemma9.1} we may assume that $|\e_j|\leq\sqrt n\log n$ for all $j=1,\ldots,n$. Then 
\begin{eqnarray*}
&&\sup_{s\in[0,1],\atop |t|>\sqrt n\log n}\left|\frac 1n\sjns \left(w_n(X_{j-1})-1\right)\left(I\{\e_j\leq t\}-F(t)\right)\right|\\
&\leq&\sup_{s\in[0,1],\atop t>\sqrt n\log n}\left|\frac 1n\sjn I\left\{\frac jn\leq s\right\} \left(w_n(X_{j-1})-1\right)\left(1-F(t)\right)\right|\\
&&{}+\sup_{s\in[0,1],\atop t<-\sqrt n\log n}\left|\frac 1n\sjn I\left\{\frac jn\leq s\right\}  \left(w_n(X_{j-1})-1\right)\left(0-F(t)\right)\right|\\
&\leq&\frac 1n\sjn \left(1-F(\sqrt n\log n)\right)+\frac 1n\sjn F(-\sqrt n\log n)\\
&=& O\left(\frac 1{n(\log n)^2}\right)\;=\;o\left(\frac 1{\sqrt n}\right),
\end{eqnarray*}
where the second last equality follows from  (\ref{9.1t^2}). 

Now let $\be=n^{-\frac 12}(\log n)^{-1}$ and let  $0=s_1<\ldots<s_{M_{n,1}}=1$ be such that $s_h-s_{h-1}\leq\be$ for all $i=2,\ldots, M_{n,1}$ and $M_{n,1}\leq 1/\be$. Further let $-\sqrt n\log n=t_1<\ldots<t_{M_{n,2}}=\sqrt n\log n$ be such that $t_i-t_{i-1}\leq\be$ for all $i=2,\ldots, M_{n,2}$ and $M_{n,2}\leq 2\sqrt n\log n/\be$. Then we have
\begin{eqnarray}
\nonumber&&\sup_{s\in[0,1],\atop t\in[-\sqrt n\log n,\sqrt n\log n]}\left|\frac 1n\sjns \left(w_n(X_{j-1})-1\right)\left(I\{\e_j\leq t\}-F(t)\right)\right|\\
\nonumber&\leq&\max_{1\leq h\leq M_{n,1},\atop 1\leq i\leq M_{n,2}}\left|\frac 1n\sum_{j=1}^{[ns_h]} \left(w_n(X_{j-1})-1\right)\left(I\{\e_j\leq t_i\}-F(t_i)\right)\right|\\
\nonumber&&+\max_{1\leq h\leq M_{n,1}}\sup_{|s-s_h|\leq\be,\atop t\in[-\sqrt n\log n,\sqrt n\log n]}\left|\frac 1n\sjns \left(w_n(X_{j-1})-1\right)\left(I\{\e_j\leq t\}-F(t)\right)\right.\\
\label{cpbewentwlemmawnj22}&&\left.\hspace{6,5cm}-\frac 1n\sum_{j=1}^{\lfloor ns_h\rfloor} \left(w_n(X_{j-1})-1\right)\left(I\{\e_j\leq t\}-F(t)\right)\right|\\
\nonumber&&+\max_{1\leq h\leq M_{n,1},\atop 1\leq i\leq M_{n,2}}\sup_{ |t-t_i|\leq\be}\left|\frac 1n\sum_{j=1}^{\lfloor ns_h\rfloor} \left(w_n(X_{j-1})-1\right)\left(I\{\e_j\leq t\}-F(t)\right)\right.\\
\label{cpbewentwlemmawnj23}&&\left.\hspace{5,5cm}-\frac 1n\sum_{j=1}^{\lfloor ns_h\rfloor} \left(w_n(X_{j-1})-1\right)\left(I\{\e_j\leq t_i\}-F(t_i)\right)\right|\hspace{2cm}\\
\label{9.3lieb}&\leq&\max_{1\leq h\leq M_{n,1},\atop 1\leq i\leq M_{n,2}}\left|\frac 1n\sum_{j=1}^{[ns_h]} \left(w_n(X_{j-1})-1\right)\left(I\{\e_j\leq t_i\}-F(t_i)\right)\right|\\
\nn&&+o(\frac{1}{\sqrt{n}})\\
\label{9.3B18}&&{}+\max_{1\leq i\leq M_{n,2}}\left|\frac 1n\sum_{j=1}^{n} \left(w_n(X_{j-1}-1)\right)\!\left(I\{\e_j\leq t_i+\be\}-F(t_i+\be)\right)\right|\\
\label{9.3B19}&&+\max_{1\leq i\leq M_{n,2}}\left|\frac 1n\sum_{j=1}^{n} \left(w_n(X_{j-1}-1)\right)\!\left(I\{\e_j\leq t_i-\be\}-F(t_i-\be)\right)\right|\\
\label{9.3mean}&&{}+2\max_{1\leq i\leq M_{n,2}}\frac 1n\sjn\left(F(t_i+\be)-F(t_i-\be)\right).
\end{eqnarray}
To obtain the last inequality it can be shown analogously to the treatment of (\ref{cpbewlemma6}) in the proof of Lemma \ref{lemma9.1} that (\ref{cpbewentwlemmawnj22}) is of order $O(\be)=o(1/\sqrt{n})$. Further 
the bounding of (\ref{cpbewentwlemmawnj23}) by the sum of (\ref{9.3B18}), (\ref{9.3B19}) and (\ref{9.3mean}) is straightforward by using monotonicity of indicator and distribution functions. 

Now by the mean value theorem and assumption (F) it follows that (\ref{9.3mean}) is of order $O(\be)=o(1/\sqrt{n})$. The remaining terms (\ref{9.3lieb}), (\ref{9.3B18}), (\ref{9.3B19})  are treated in the same way and we will only consider (\ref{9.3lieb}) in what follows. For this term we have for each $\eta>0$ that
\begin{eqnarray*}
&&P\left(\sqrt n\max_{1\leq h\leq M_{n,1},\atop 1\leq i\leq M_{n,2}}\left|\frac 1n\sum_{j=1}^{\lfloor ns_h\rfloor} \left(w_n(X_{j-1})-1\right)\left(I\{\e_j\leq t_i\}-F(t_i)\right)\right|>\eta\right)\\
&\leq&\sum_{h=1}^{M_{n,1}}\sum_{i=1}^{M_{n,2}}P\left(\sqrt n\cdot\left|\frac 1n\sum_{j=1}^{\lfloor ns_h\rfloor} \left(w_n(X_{j-1})-1\right)\left(I\{\e_j\leq t_i\}-F(t_i)\right)\right|>\eta\right)\\
&\leq&M_{n,1}M_{n,2}\left(4\exp\left(-\frac{n\eta^2}{64n\omega_n+\frac 83  \sqrt n\eta \lfloor n^{\frac 12}(\log n)^{-2}\rfloor}\right)+4\frac n{\lfloor n^{\frac 12}(\log n)^{-2}\rfloor}\alpha(\lfloor n^{\frac 12}(\log n)^{-2}\rfloor)\right) \\
&=& o(1)
\end{eqnarray*}
(where $\alpha(\cdot)$ denotes the $\alpha$-mixing coefficient)
by an application of Theorem 2.1 by Liebscher (1996) and the  bandwidth conditions. Details are omitted for the sake of brevity, but note that
\beq
\omega_n&=&\frac 1{\lfloor n^{\frac 12}(\log n)^{-2}\rfloor}\max_{0\leq S\leq n-\lfloor n^{\frac 12}(\log n)^{-2}\rfloor}\sum_{j=S+1}^{S+\lfloor n^{\frac 12}(\log n)^{-2}\rfloor}E\left[(w_n(X_{j-1})-1)^2\right]\\
&\leq&\int_{-\infty}^{a_n+\kappa}f_{X_0}(x)dx+\int_{b_n-\kappa}^{\infty}f_{X_0}(x)dx\;=\;o\left(\frac 1{\log n}\right)
\eeq
by assumption (I). 
\hfill $\Box$

\medskip

\noindent {\bf Proof of Lemma \ref{lemma9.4}.}  
We only give arguments for the first statement. Similarly to the proof of Lemma \ref{lemma9.3} one can show that
\beq
\frac{1}{n}\sum_{j=1}^{\lfloor n\theta_0\rfloor} w_{nj} \left(I\left\{{\e}_j\leq t\right\}-F(t)\right)
&=& \frac{1}{n}\sum_{j=1}^{\lfloor n\theta_0\rfloor} \left(I\left\{{\e}_j\leq t\right\}-F(t)\right)+o_P(1)
\eeq
(applying assumptions (I') and (X')). 
Then the assertion follows by standard arguments for the empirical distribution function of iid-data. 
\hfill $\Box$



\section*{References}

\begin{description}

\item Andreou, E. \& Ghysels, E. (2009) \textit{Structural Breaks in Financial Time Series.} T. G. Andersen (ed) et al, Handbook of Financial Time Series. Springer, Berlin, 839-870.
 
\item  Bai, J. (1994). \textit{Weak convergence of the sequential empirical processes of residuals in ARMA models.} Ann. Statist. 22, 2051-2061.
 
  \item Boldin, M. V. (2002).  \textit{On sequential residual empirical processes in heteroscedastic time series.}  Math. Methods Statist. 11, 453-464.

\item Cs\"org\"o, M., Horvßth, L. \& Szyszkowicz (1997). \textit{Integral tests for suprema of Kiefer processes with application}. Statist. Decisions 15, 365-377.

 \item Dette, H., Pardo-Fernßndez, J. C. \& Van Keilegom, I. (2009). \textit{Goodness-of-Fit Tests for Multiplicative Models with Dependent Data}. Scand. J. Statist. 36, 782-799.

\item Doukhan, P. (1994). \textit{Mixing, Properties and Examples}. Springer, New York.

\item Fan, J. \& Yao, Q. (2005). \textit{Nonlinear Time Series}. Springer, New York.

\item Freedman, D. A. (1975). \textit{On tail probabilities for martingals}. Ann. Probab. 3, 100-118.
 
 \item Giraitis, L., Leipus, R. \& Surgailis, D. (1996) \textit{The change-point problem for dependent observations.} J. Statist. Plann. Inf. 53, 297-310.
 
 \item Hansen, B. E. (2008). \textit{Uniform Convergence Rates for Kernel Estimation with Dependent Data}. Econom. Theory 24, 726-748.
 
\item Hlßvka, Z., Hu\v{s}kovß, M., Kirch, C. \& Meintanis, S. (2012). \textit{Monitoring changes in the error distribution of autoregressive models based on Fourier methods}. to appear in Test.
 
 \item Horv\'ath, L., Kokoszka, P. \& TeyssiÞre, G. (2001). \textit{Empirical process of the squared residuals of an ARCH sequence.} Ann. Statist. 29, 445-469.
 
 \item Hu\v{s}kovß, M. \& Antoch, J. (2003). \textit{Detection of structural changes in regression.} Tatra Mt. Math. Publ. 26, 201-215.
  
 \item Hu\v{s}kovß, M., Prß\v{s}kovß, Z. \& Steinebach, J. (2007). \textit{On the detection of changes in autoregressive time series. I. Asymptotics. } J. Statist. Plann. Inference 137, 1243-1259.
 
  \item Hu\v{s}kovß, M., Kirch, C., Prß\v{s}kovß, Z. \& Steinebach, J. (2008). \textit{On the detection of changes in autoregressive time series. II. Resampling. } J. Statist. Plann. Inference 138, 1697-1721.

\item Inoue, A. (2001). \textit{Testing for distributional change in time series.} Economet. Theory 17, 156-187.

 \item Kirch, C. \& Tadjuidje Kamgaing, J. (2012). \textit{Testing for parameter stability in nonlinear autoregressive models}. J. Time Ser. Anal., to appear.

 \item Koul, H. L. (1996). \textit{Asymptotics of some estimators and sequential residual empiricals in nonlinear time series.} Ann. Statist. 24, 380-404.
 
 \item Koul, H. L. (2002). \textit{Weighted Empirical Processes in Dynamic Nonlinear Models (Second Edition).} Springer, New York.

\item Krei▀, J.-P. (1991). \textit{Estimation of the distribution function of noise in stationary processes.} Metrika 38, 285-297.

\item Lee, S. \&  Na, S. (2004). \textit{A nonparametric test for the change in the density function in strong mixing processes.}  Statist. Prob. Letters 66, 1-25.

\item Liebscher, E. (1996). \textit{Strong convergence of sums of $\alpha$-mixing random variables with applications to density estimation}. Stochastic Processes and their Applications 65, 69-80.
 
\item M³ller, U. U., Schick, A. \& Wefelmeyer, W. (2009). \textit{Estimating the innovation distribution in nonparametric autoregression}. Probab. Theory Relat. Fields 144, 53-77.

\item Neumeyer, N. \& Van Keilegom, I. (2009). \textit{Change-Point Tests for the Error Distribution in Nonparametric Regression}. Scand. J. Statist. 36, 518-541.

\item Picard, D. (1985). \textit{Testing and estimating change-points in time series}. Adv. Appl. Probab. 17, 841-867.

\item Pollard, D. (1990). \textit{Empirical Processes: Theory and Applications}. NSF-CBMS Regional Conference Series in Probability and Statistics 2, Institute of Mathematical Statistics.
 
\item Selk, L. (2011). \textit{Change-Point-Tests f³r die Innovationenverteilung in nichtparametrischen Autoregressionsmodellen auf Basis sequentieller empirischer Prozesse}. PhD thesis (in German), Universitõt Hamburg. http://ediss.sub.uni-hamburg.de/volltexte/2011/5338/

 \item Shao, X. \& Zhang, X. (2010). \textit{Testing for Change Points in Time Series}. J. Amer. Statist. Assoc. 105, 1228-1240.

\item Shorack, G. R. \&  Wellner, J.\,A. (1986). \textit{Empirical Processes with Applications ot Statistics.} Wiley, New York. 
 
 \item Shumway, R. H. \& Stoffer, D.S. (2006). \textit{Time Series Analysis and Its Applications: With R Examples}. Springer, New York.
 
 \item van der Vaart, A. W. \& Wellner, J. A. (1996). \textit{Weak Convergence and Empirical Processes}. Springer, New York.
 
 \end{description}

\end{appendix}

\end{document}